\documentclass[
 amsmath,amssymb,
 aps, prapplied, reprint, superscriptaddress, longbibliography,
twocolumn
]{revtex4-2}
\usepackage{graphicx}
\usepackage{tikz}

\newcommand{\bra}[1]{\left\langle#1\right|}
\newcommand{\ket}[1]{\left|#1\right\rangle}
\newcommand{\matrixel}[3]{\left\langle #1 \middle| #2 \middle| #3 \right\rangle}

\begin{document}

\title{Reduced leakage in single-qubit transmon gates driven by single flux quantum pulses}

\author{Lucas A. Rogers}
\affiliation{Department of Physics, University of
Wisconsin-Madison, Madison, Wisconsin 53706, USA}

\author{Oleg Mukhanov}
\affiliation{SEEQC, Inc., Elmsford, New York 10523, USA}

\author{Maxim G. Vavilov}
\affiliation{Department of Physics, University of
Wisconsin-Madison, Madison, Wisconsin 53706, USA}

\date{\today}

\begin{abstract}

We propose a design for single flux quantum (SFQ) control sequences for transmon qubits that effects fast single-qubit gates via geometric leakage suppression. We apply a sequence composed of an integer number of subsequences of two or more SFQ pulses, or kicks, applied approximately over the qubit period and spaced by $T_{12}$ divided by the pulse count, where $T_{12}$ is the transition period between the first and second excited states. This novel geometrically-motivated subsequence structure cancels the lowest-order transition amplitude into the second excited state. At zero offset charge this results in the single-period leakage into the second excited state scaling quartically in the SFQ pulse strength, while the computational rotation accumulates linearly. This permits strong kicks and thus allows for fast gates with low coherent error without the engineering overhead associated with pulse shaping or complex on-chip electronics. For $X$ gates implemented on a representative transmon, we achieve coherent gate error below $10^{-4}$ at gate times $< 15 \text { ns}$, with $X/2$ gates achieving error below $ 10^{-4}$ at $4.3 - 6.3 \text{ ns}$. Conventional single-kick sequences require 50 ns to reach $10^{-3}$ and do not reach $10^{-4}$ within 130 ns. We calculate the detuning required for the subsequence repetition period relative to the qubit period, which is proportional to the square of the kick strength, and consider the sensitivity of the gate relative to offset charge noise, clock-frequency mismatch, timing jitter and pulse loss. Total coherent error is brought into the vicinity of $10^{-7}$ with ideal sequences, which is far below the decoherence threshold with practical devices.

\end{abstract}

\maketitle

\section{Introduction}

Superconducting quantum processors with fault-tolerant qubit counts place substantial demands on the control electronics. Conventional superconducting qubit control is implemented via arbitrary waveform generators that synthesize their microwave pulses with room-temperature electronics and send them into the cryostat via heavily attenuated coaxial lines, which becomes geometrically and thermally difficult to scale to high qubit counts~\cite{Jordan2026,kawabata2026integration-9f2,Cunningham2025Breathers,Xia2026Subharmonic}. Single flux quantum (SFQ) logic allows the control circuits themselves to be operated at cryogenic temperatures, generating digital sequences of identical voltage pulses that drive the qubit effectively without introducing the thermal and packaging problems associated with the usage of classical control lines \cite{mcdermott1,mcdermott2014accurate,Leonard_2019,Liu_2023,Weng2023SFQ,vozhakov2023speeding-bc0,Vozhakov2026DRAGSFQ,Sun2026DigitalControl}.

The tradeoff with SFQ versus classical control is reduced freedom in drive shape. Contrary to the analog nature of classical shaped microwave pulse control, an SFQ pulse is approximately equivalent to a delta-function `kick' of fixed, quantized area, and so control sequences are created by varying the quantity of and timing between kicks, with kick strength adjustment generally fixed at fabrication~\cite{liu2023single-flux-quantum-based-31c} by choosing the coupling capacitance. The simplest realization of SFQ control applies one kick per qubit Larmor period \cite{mcdermott2014accurate,Leonard_2019}. For transmons \cite{Koch_2007}, whose $\ket{1} \rightarrow \ket{2}$ transition period $T_{12}$ is close to the computational transition period $T_{01}$, this near-resonance can result in substantial accumulation of population in $\ket{2}$. This leakage from the computational subspace sets the practical limit on how strongly the qubit can be driven and thus on how fast the gate operation can be completed. Optimization against leakage also motivated the development of derivative-based pulse shaping for microwave control \cite{Motzoi_2009,Hyyppa2024FASTDRAG}. For SFQ control, numerically optimized sequences, including those built from repeated short subsequences, have been shown to reduce leakage \cite{Li2019HardwareEfficient, mcdermott1,Liebermann_2016,shillito2023compact-6a0,vozhakov2023speeding-bc0,Vozhakov2026DRAGSFQ} but can require gate- and qubit-specific pulse schedules, substantially complicating cryogenic control electronics in terms of power dissipation and integration scale per qubit.

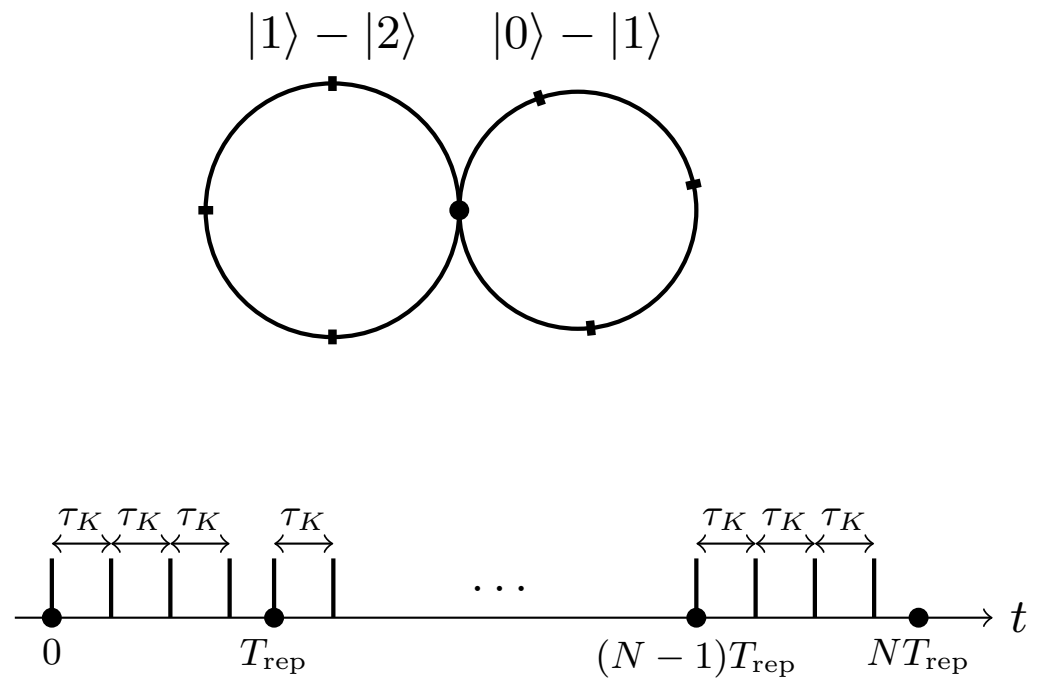
\begin{figure}[htbp]
\centering
\begin{tikzpicture}[scale=0.49]
    \path[use as bounding box] (-6.6,-6.8) rectangle (6.6,2.7);
    \def\Rtwelve{1.712}
    \def\Rzeroone{1.6}
    \begin{scope}[shift={(-\Rtwelve,0)}]
        \draw[thick] (0,0) circle (\Rtwelve);
        \node at (0, 2.35) {$|1\rangle - |2\rangle$};
        \foreach \i in {1,2,3} {
            \pgfmathsetmacro{\ang}{\i*90}
            \draw[ultra thick] ({\ang}:1.61) -- ({\ang}:1.81);
        }
    \end{scope}
    \begin{scope}[shift={(\Rzeroone,0)}]
        \draw[thick] (0,0) circle (\Rzeroone);
        \node at (0, 2.35) {$|0\rangle - |1\rangle$};
        \foreach \i in {1,2,3} {
            \pgfmathsetmacro{\ang}{180 + \i*96.3}
            \draw[ultra thick] ({\ang}:1.5) -- ({\ang}:1.7);
        }
    \end{scope}
    \draw[fill=black] (0,0) circle (0.12);

    \begin{scope}[shift={(0,-5.5)}]
        \draw[->] (-6,0) -- (7.2,0) node[right] {$t$};

        \foreach \i/\x in {0/0, 1/0.8, 2/1.6, 3/2.4} {
            \draw[thick] (-5.5+\x, 0) -- (-5.5+\x, 0.8);
            \ifnum \i=0
                \draw[fill=black] (-5.5+\x, 0) circle (0.12);
            \fi
            \ifnum \i<3
                \pgfmathsetmacro{\nextx}{\x + 0.8}
                \draw[<->, thin] (-5.5+\x, 1.0) -- (-5.5+\nextx, 1.0);
                \node[above, scale=0.8] at (-5.5+\x+0.4, 1.0) {$\tau_K$};
            \fi
        }
        \node[below, scale=0.8] at (-5.5, -0.1) {$0$};

        \pgfmathsetmacro{\TzeroOnePos}{-5.5 + 3.0}
        \foreach \i/\x in {0/0, 1/0.8} {
            \draw[thick] (\TzeroOnePos+\x, 0) -- (\TzeroOnePos+\x, 0.8);
            \ifnum \i=0
                \node[below, scale=0.8] at (\TzeroOnePos, -0.1) {$T_{\rm rep}$};
                \draw[fill=black] (\TzeroOnePos+\x, 0) circle (0.12);
                \draw[<->, thin] (\TzeroOnePos, 1.0) -- (\TzeroOnePos+0.8, 1.0);
                \node[above, scale=0.8] at (\TzeroOnePos+0.4, 1.0) {$\tau_K$};
            \fi
        }

        \node at (0.6, 0.4) {$\dots$};

        \pgfmathsetmacro{\finalStart}{3.2}
        \foreach \i/\x in {0/0, 1/0.8, 2/1.6, 3/2.4} {
            \draw[thick] (\finalStart+\x, 0) -- (\finalStart+\x, 0.8);
            \ifnum \i=0
                \draw[fill=black] (\finalStart+\x, 0) circle (0.12);
            \fi
            \ifnum \i<3
                \pgfmathsetmacro{\nextx}{\x + 0.8}
                \draw[<->, thin] (\finalStart+\x, 1.0) -- (\finalStart+\nextx, 1.0);
                \node[above, scale=0.8] at (\finalStart+\x+0.4, 1.0) {$\tau_K$};
            \fi
        }
        \node[below, scale=0.8] at (\finalStart, -0.1) {$(N-1)T_{\rm rep}$};
        \draw[fill=black] (\finalStart+3.0, 0) circle (0.12);
        \node[below, scale=0.8] at (\finalStart+3.0, -0.1) {$NT_{\rm rep}$};
    \end{scope}
\end{tikzpicture}
\caption{Geometric leakage suppression (top) and the periodic pulse train (bottom), drawn for $K = 4$. Each circle is one transition period, with a radius proportional to it, so that equal arcs are equal times. The circles touch at the bold dot, which represents the first kick of a period. The spacing $\tau_K = T_{12}/K$ causes the first-order $\ket{1}\rightarrow\ket{2}$ amplitudes to cancel, while over the shorter $\ket{0} \leftrightarrow \ket{1}$ period the same kicks do not completely destructively interfere and the rotation accumulates as the structure repeats with period $T_{\rm rep} = (1+\alpha)T_{01} \approx T_{01}$ [Eq.~\eqref{eq:trep}]. Bold dots in the `comb' diagram mark the beginning of each period and the gate completion.}
\label{fig:pulse_train}
\end{figure}

In this paper we consider periodic sequences that are instead specified by three numerical parameters and a period counter, which is substantially less complex and compatible with implementation in the cryogenic portion of the device. We show that the tradeoff between leakage and gate time can be significantly ameliorated by manipulating the timing of the per-period pulse sequence, which we refer to as the subsequence. We distribute $K\geq2$ SFQ kicks with spacing $\tau_K = T_{\text{12}} / K$ within each qubit Larmor period, so that the first-order $\ket{1}\rightarrow\ket{2}$ amplitudes corresponding to each kick acquire equal phase increments over a full period on the $\ket{1}, \ket{2}$ Bloch sphere and interfere destructively. However, the destructive interference in the computational subspace is incomplete because $T_{01} \neq T_{12}$ (Fig.~\ref{fig:pulse_train}). At zero offset charge, this technique of geometric cancellation suppresses single-period leakage to $\ket{2}$ to quartic order in the pulse strength, which allows the use of stronger pulses and thus faster gates with minimal leakage penalty.

The subsequence is specified by the following three parameters: $K$, the kick count; $\delta \theta$, the kick strength; and the compensatory detuning of the subsequence period from the Larmor period in response to the frequency shifts induced by the kicks themselves, which we denote $\alpha$. The choice of $K$ involves a tradeoff between the performance of the gate and the peak pulse rate required from the SFQ controller. (For the transmon considered here, $K = 4$ represents a favorable compromise, as discussed in Sec.~\ref{sec:results}.)  With superconducting qubits such as transmons, any deviation from the intended `longitude' ($\phi$) after the sequence can be corrected at the end with a virtual $Z$ gate, performed in software and costing no gate time \cite{McKay_2017}.

The paper is organized as follows: in Section~\ref{sec:model}, we introduce the transmon-SFQ model and demonstrate the construction of the subsequence, sequence and their evolution operators. In Section~\ref{sec:single_period_leakage}, we establish the geometric suppression of leakage from the computational subspace, which is the central result of this work, using a perturbative expansion confirmed by exact numerical simulation. In Section~\ref{sec:results}, we describe the calibration of the sequence and present gate times and errors confirmed numerically. Subsequent sections discuss the robustness of this protocol to the most common error channels: in Section~\ref{sec:charge_noise}, we consider offset charge or clock drift; and in Section~\ref{sec:jitter}, timing jitter of the SFQ pulse train (stochastic pulse drops are discussed in Appendix~\ref{app:pulse_drops}). Finally we conclude by summarizing the gate performance and discussing implications for cryogenic qubit control.

\section{System and SFQ drive}
\label{sec:model}

\subsection{The SFQ-driven transmon}

Transmons are superconducting qubits designed to operate in the regime where the Josephson energy $E_J$ significantly exceeds the charging energy $E_C$, rendering them insensitive to charge noise \cite{Koch_2007,Krantz2019QuantumEngineer}. The transmon Hamiltonian is
\begin{equation}
    \hat{H} = 4E_C (\hat{n} - n_g)^2 - E_J \cos \hat{\phi},
    \label{eq:transmon_H}
\end{equation}
where $\hat{n} = \hat{Q} / 2e$ is the Cooper pair number operator and $\hat{\phi}$ is the phase operator, with $[ \hat{n}, \hat{\phi} ] = -i$, and $n_g$ is the offset charge. We have $\omega_{01} \approx (\sqrt{8 E_J E_C} - E_C) / \hbar$ and $\omega_{12} - \omega_{01} \approx - E_C / \hbar$. Throughout, $\omega_{ij} = (E_j - E_i)/\hbar$ is a transition \textit{angular} frequency and $f_{ij} = \omega_{ij} / 2 \pi$ is a transition frequency in cycles per unit time; we use whichever is more convenient and quote numbers as $f_{ij}$ in GHz.

The primary characteristic of the transmon for quantum control is its weak negative anharmonicity. We define the Larmor transition periods as
\begin{align}
    T_{01} &= 2 \pi / (\omega_1 - \omega_0), \nonumber \\
    T_{12} &= 2 \pi / (\omega_2 - \omega_1) \text{ with } T_{12} > T_{01}.
\end{align}
This small disparity between $T_{01}$ and $T_{12}$ leads to a strong propensity for leakage to $\ket{2}$, which is precisely what our control strategy must accommodate.

The SFQ drive consists of a sequence of timed kicks: voltage pulses for which the time integral is equal to a single magnetic flux quantum $\Phi_0 = h / 2e$. A voltage $V_g(t)$ applied over time to the qubit's island through a coupling capacitor with capacitance $C_C$ alters the Hamiltonian via shifting the gate offset charge:
\begin{equation}
    n_{g, \, \mathrm{SFQ}} = n_g + \frac{C_C}{2e} V_g(t),
    \label{eq:gatechargeshift}
\end{equation}
which then gives the drive term (with $C_\Sigma$ being the total capacitance)
\begin{equation}
    \hat{H}_{\mathrm{SFQ}} = -2e \frac{C_C}{C_\Sigma} V_g (t) \left( \hat{n} - n_g \right).
    \label{eq:sfqhamiltonian}
\end{equation}
Then we introduce the dimensionless quantity $\delta \theta$, the kick strength, which is equal to
\begin{equation}
    \delta \theta = \frac{2e C_C}{\hbar C_{\Sigma}} \int V_g(t) \, dt = 2 \pi \frac{C_C}{C_\Sigma},
    \label{eq:dtheta}
\end{equation}
and with the SFQ voltage pulse $V_g (t)$ being treated as having zero duration, we recover the following equation for the SFQ pulse operator. We should note that this approximation requires a short pulse duration: in Appendix~\ref{app:width} we demonstrate numerically that the approximation is valid for realistic pulse widths of around 2 ps.
\begin{equation}
    \hat{U}_{\mathrm{SFQ}} (\delta \theta) = e^{i \delta \theta (\hat{n} - n_g)}.
    \label{eq:u_sfq}
\end{equation}
We will drop $n_g$ from $\hat{U}_{\mathrm{SFQ}}$ after this point because it is an irrelevant global phase, and note for clarity that the charge operator $\hat{n}$ is represented in the transmon Hamiltonian eigenbasis. We will also note that the sign of the exponent here is opposite to that of the drive term $\hat{H}_{\mathrm{SFQ}}$ and that a pulse of opposite polarity simply inverts the offset charge and changes none of the following results, aside from reflecting the offset-charge dependence discussed in Sec.~\ref{sec:charge_noise}. Representative kick strengths of $\delta \theta \in [0.01, 0.05]$ correspond to values of $C_C / C_\Sigma$ between approximately $1.6 \times 10^{-3}$ and $8 \times 10^{-3}$. The wider range of kick strengths that we consider in numerical calculations is discussed in Sec.~\ref{sec:results}.

\subsection{Subsequence construction}

The gate is assembled in subsequences of $K$ kicks, the repetition of which creates a sequence of $N$ subsequences. Between the kicks are intervals of free evolution:
\begin{equation}
    \hat{U}_{\mathrm{free}}(t) = e^{-i\hat{H}t / \hbar},
    \label{eq:u_free}
\end{equation}
which is diagonal in the eigenstate basis, with $[\hat{U}_{\mathrm{free}}(t)]_{nn} = e^{-i \omega_n t}$.

With a subsequence of $K$ kicks applied over one repetition period $T_{\mathrm{rep}} \approx T_{01}$ [Eq.~\eqref{eq:trep}], we implement suppression of leakage to $\ket{2}$ by distributing the kicks with equal temporal separation $\tau_K = T_{12} / K$.

The physical intuition behind this strategy is captured by the time circles in Fig.~\ref{fig:pulse_train}. In the $\ket{1}, \ket{2}$ subspace, the equal intervals between pulses on the period $T_{12}$ are represented by the equal spacing of the dots representing them around the time circle. It can be said that the center of mass of the dots is at the center of the circle, which is representative of first-order cancellation of the $\ket{1}\rightarrow\ket{2}$ transition amplitude.

Equivalently, we can argue that for obvious physical reasons the absolute values of the transition amplitudes for each kick, $|a_\mathrm{kick}|$, will be equal. From the timing they all acquire a relative phase of $e^{i 2 \pi / K}$ within the $\ket{1}, \ket{2}$ subspace, and as those are complex roots of unity, we have at first order in $\delta \theta$, for $K \geq 2$,
\begin{equation}
    A_{21}^{(1)} = i \delta \theta \matrixel{2}{\hat{n}}{1} \sum_{m=0}^{K-1} e^{i 2 \pi m / K} = 0.
    \label{eq:phase_cancellation}
\end{equation}
This demonstrates that under the application of this subsequence, the first order transition amplitude from $\ket{1} \rightarrow \ket{2}$ vanishes. Accepting this argument, we can see that the center of mass of the dots on the $\ket{0}, \ket{1}$ time circle is off-center, meaning that $\sum_{k=0}^{K-1} e^{i\omega_{01} k\tau_K} \neq 0$.

To build a sequence, and thus an $X$ gate, from this, we repeat the subsequence $N$ times over a repetition period $T_{\mathrm{rep}}$ very close to $T_{01}$:
\begin{equation}
    T_{\text{rep}} = (1 + \alpha)\, T_{01}, \qquad |\alpha| \ll 1,
    \label{eq:trep}
\end{equation}
where $\alpha$ is a detuning that, for calibrated $X$ gates, scales as $(\delta \theta)^2$ at weak kick strengths to compensate for the shift in the qubit Larmor period induced by the pulses themselves. Defining $\Delta T = T_{12} - T_{01} > 0$ for convenience, we can construct the operator for one repetition period:
\begin{multline}
    \hat{U}_{\text{per}} = \hat{U}_{\text{free}}(\tau_K - \Delta T + \alpha T_{01}) \\
    \times \left[ \hat{U}_{\text{SFQ}}(\delta\theta)\, \hat{U}_{\text{free}}(\tau_K) \right]^{K-1} \hat{U}_{\text{SFQ}}(\delta\theta).
    \label{eq:u_period}
\end{multline}
Note that the final free-evolution interval is

\begin{equation}
    \tau_{\mathrm{last}} = T_{\mathrm{rep}} - (K-1) \tau_K = \tau_K - \Delta T + \alpha T_{01}.
    \label{eq:tau_last}
\end{equation}
This ensures that the total subsequence duration is equal to $T_{\mathrm{rep}}$. Requiring $\tau_{\mathrm{last}}>0$, and assuming $\Delta T-\alpha T_{01}>0$, gives
\begin{equation}
    K < \frac{T_{12}}{\Delta T - \alpha T_{01}} \approx \frac{1}{1 - T_{01}/T_{12}}.
    \label{eq:positive_endrun_condition}
\end{equation}
The last form is the $\alpha \to 0$ limit. For example, with the parameters used below $(E_J/E_C = 75, T_{12}/T_{01} = 1.05)$, this condition is satisfied for $K \lesssim 21$.

The gate operator $\hat{U}_{\mathrm{gate}}$ is then
\begin{equation}
    \hat{U}_{\text{gate}} = (\hat{U}_{\text{per}})^N, \qquad T_{\text{gate}} = N\, T_{\text{rep}}.
    \label{eq:u_total}
\end{equation}
The single-period \textit{subsequence} repeated $N$ times constitutes the \textit{sequence}, which implements the gate. It can now be noted that, since a given rotation in the computational subspace is effected by the action of $\hat{U}_{\mathrm{per}}$, $N$ depends on $K$, $\delta\theta$, $\alpha$ and the target gate. The detuning $\alpha$ is determined by those parameters as well for the $X$-gate, with further discussion in App.~\ref{app:alpha}. For the $X/2$-gate, the value of $\alpha $ is free over the window $2 \pi | \alpha - \alpha_X | \lesssim \Theta$, where $\Theta$ is the angle of the transverse computational rotation effected by $\hat{U}_{\text{per}}$, given to leading order by Eq.~\eqref{eq:bigtheta}. Here $\alpha_X$ is the calibrated $X$-gate detuning.

While this effects the desired rotation around an axis in the equatorial plane of the Bloch sphere for the $X$ gate, it does not specify the resulting longitude of the statevector after the gate. Fortunately, with superconducting qubits, this can be corrected by a virtual $Z$ rotation implemented in software. Even if implemented by shifting the subsequent pulse train, this results in a delay of less than one qubit period.

It is sensible to briefly discuss virtual-$Z$ error. When implementing the virtual-$Z$ rotation via shifting the following pulse train, a timing error $\Delta t $ corresponds to a phase error $\Delta \phi = \omega_{01} \Delta t$. This results in an additional gate infidelity of approximately

\begin{equation}
    (1-F)_Z \simeq \frac{(\omega_{01} \Delta t)^2}{6} = \frac{(\Delta \phi)^2}{6}.
    \label{eq:virtual_z_error}
\end{equation}
Thus, for the qubit frequency used below of 4.690 GHz, the virtual-$Z$ timing error must be less than approximately 0.83 ps so that $(1-F)_Z < 10^{-4}$.

\section{Single-period leakage analysis}
\label{sec:single_period_leakage}

Leakage is the process of population of states outside the computational subspace. We will define the leakage into state $n$ after the application of an arbitrary unitary operator $\hat{U}$ as:
\begin{equation}
    L_n = \frac{1}{2} \left( |\langle n | \hat{U} | 0 \rangle|^2 + |\langle n | \hat{U} | 1 \rangle|^2 \right).
    \label{eq:leakage_def}
\end{equation}
In Section~\ref{sec:numerics}, we compute the $L_n$ numerically. Here we will analyze the leakage perturbatively in powers of $\delta \theta$ and explain the physical origin of our technique's leakage suppression.

\begin{figure*}[t]
\centering
\includegraphics[width=0.86\textwidth]{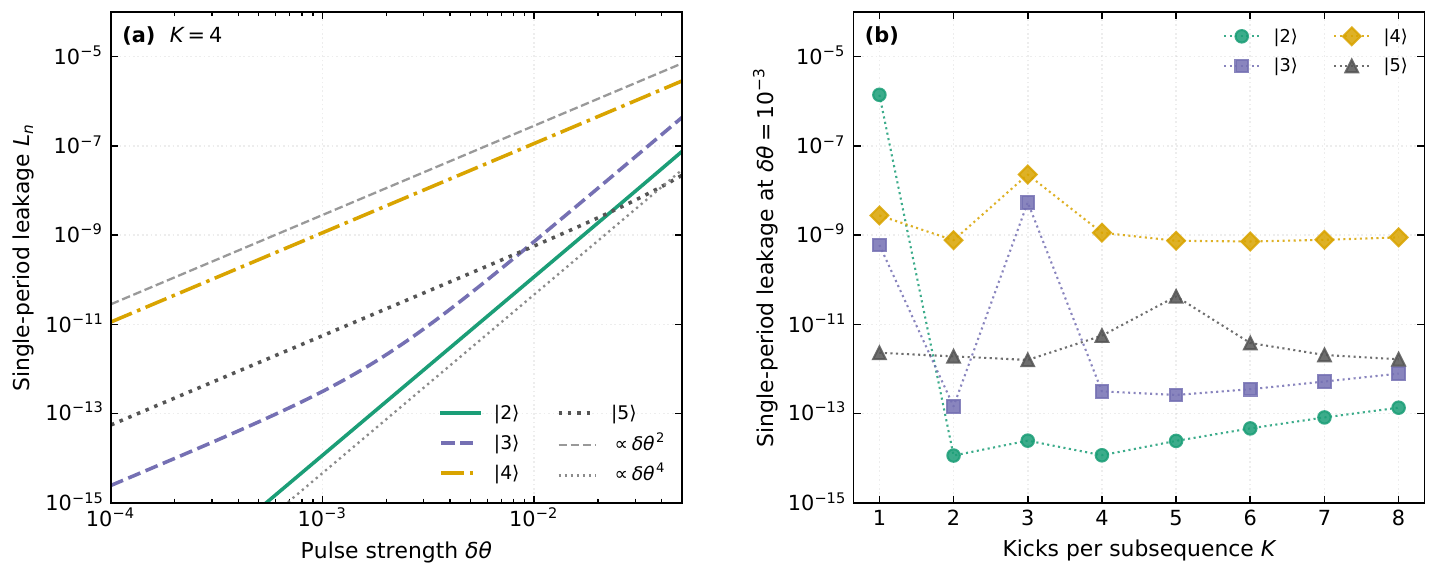}
\caption{Single period leakage $L_n$ for $\ket{n} = \ket{2} \text{ through } \ket{5}$ with $E_C/h = 0.2$ GHz and $E_J / h = 15$ GHz. (a) $L_n$ is plotted against $\delta \theta$ for $K = 4$ on a log-log scale. $\ket{2}$ exhibits a slope of 4, indicating suppression to quartic order in the pulse strength, whereas $\ket{3}$, $\ket{4}$ and $\ket{5}$ display a slope of 2, scaling with $\delta \theta ^2$ at low amplitudes. The bend in the plot for $\ket{3}$ arises because the first-order amplitude is nearly suppressed by the resonance $\omega_{03} \approx 3 \omega_{12}$ [Eq.~\eqref{eq:comb}], whereas after $\delta \theta \approx 2 \times 10^{-3}$ the second-order amplitude takes over. (b) $L_n$ is plotted versus the kick number $K$ at fixed $\delta \theta = 10^{-3}$. Going from $K = 1$ to $K = 2$ causes $L_2$ to fall by eight orders of magnitude. $L_3$ is strongly suppressed for every value of $K$ not dividing 3 due to the aforementioned resonance condition between $\omega_{03}$ and $\omega_{12}$. Otherwise, all leakage channels are nearly independent of $K$.}
\label{fig:leakage_K8}
\end{figure*}

\subsection{Perturbative analysis of leakage behavior}
\label{sec:perturbation}

In order to understand the scaling of the leakage into $\ket{3}$ and $\ket{4}$, we will expand the eigenstates $\ket{n}$ in terms of the harmonic oscillator eigenstates $\ket{n}^{(0)}$ using first-order perturbation theory. We will treat the transmon as a weakly anharmonic Duffing oscillator, which means that the leading-order correction to the Hamiltonian is the quartic term given in $\hat{H}_1$:
\begin{subequations}
\begin{align}
    \hat{H}_0 = \hbar \omega_{p} \left( \hat{a}^\dagger \hat{a} + \frac{1}{2} \right)&, \\
    \hat{H}_1 = -\frac{E_C}{12} \left( \hat{a} + \hat{a}^\dagger \right)^4&
\end{align}
\end{subequations}
with $\omega_p = \sqrt{8 E_J E_C} / \hbar$ being the qubit plasma frequency. We can represent the Cooper pair number operator $\hat{n}$, relevant because $\hat{U}_{\mathrm{SFQ}} = e^{i \delta \theta \hat{n}}$, in terms of the bosonic creation and annihilation operators as well as the zero-point charge fluctuation amplitude (in Cooper pair number units), $n_0$:
\begin{equation}
    \hat{n} = i n_0 \left( \hat{a}^\dagger - \hat{a} \right), \qquad n_0 = \left( \frac{E_J}{32 E_C} \right)^{1/4}.
\end{equation}
In the harmonic oscillator limit, we have the usual selection rules dictating that the Cooper pair number operator only couples adjacent states. Introducing the quartic perturbation $\hat{H}_1$ allows mixing of same-parity states via the $( \hat{a} + \hat{a}^\dagger )^4$ term.

Applied to the $\ket{0} \rightarrow \ket{3}$ and $\ket{1} \rightarrow \ket{4}$ transitions, the selection rules restrict the sets of intermediate states to $\{ \ket{1}, \ket{2}, \ket{4} \}$ and $\{ \ket{0}, \ket{2}, \ket{3}, \ket{5} \}$ respectively. We use first-order perturbation theory to recover the following matrix elements, with $\lambda = \sqrt{E_C / (8 E_J)} = E_C / \hbar \omega_p$ the dimensionless perturbation parameter, and with $n_g=0$ throughout this section:
\begin{subequations}
\label{eq:30and41amps}
\begin{align}
    |\langle 3 | \hat{n} | 0 \rangle| &\approx n_0 \lambda \frac{\sqrt{6}}{4}, \label{eq:30amp} \\
    |\langle 4 | \hat{n} | 1 \rangle| &\approx n_0 \lambda \frac{\sqrt{6}}{2}.
    \label{eq:41amp}
\end{align}
\end{subequations}
This result reveals that the transmon's weak anharmonicity means that the above matrix elements are allowed at first order in the state mixing, and that the $\ket{1} \rightarrow \ket{4}$ coupling is twice as strong as that for $\ket{0} \rightarrow \ket{3}$. Exact diagonalization at $E_J/E_C = 75$ ($\lambda = 0.041$, $n_0 = 1.237$) gives $|\langle 3 | \hat{n} | 0 \rangle| = 0.0347$ and $|\langle 4 | \hat{n} | 1 \rangle| = 0.0743$, a ratio of $2.14$, against first-order estimates of $0.0309$ and $0.0619$.

For a small value of $\delta \theta \ll 1$ we can expand the SFQ operator $\hat{U}_{\text{SFQ}} = e^{i \delta \theta \hat{n}} \approx \hat{I} + i \delta \theta \hat{n}$.

Over a complete subsequence, the first-order transition amplitudes must account for the phases accumulated between kicks. For a given transition $\ket{i} \rightarrow \ket{f}$, these phases are accounted for with the interference or comb factor
\begin{equation}
    S_K(\omega ) = \sum_{k=0}^{K-1} e^{i \omega k \tau_K } = \frac{\sin (K \omega \tau_K / 2)}{\sin (\omega \tau_K / 2)} e^{i (K-1) \omega \tau_K /2}.
    \label{eq:comb}
\end{equation}
Here $\omega$ is the angular frequency of the transition. After one complete subsequence, the transition amplitude to first order in $\delta \theta$ is proportional to $\delta \theta \, n_{fi} S_K(\omega_{if})$ (up to an overall phase).

As the charge matrix elements $n_{30}, n_{41} \neq 0$, the leading order leakages into $\ket{3}$ and $\ket{4}$ after a single subsequence are thus
\begin{subequations}
\label{eq:L3andL4}
\begin{align}
    L_3(\delta\theta) &\approx \frac{1}{2} \delta\theta^2 |n_{30}|^2 |S_K (\omega_{03})|^2 \propto \delta\theta^2, \label{eq:L3} \\
    L_4(\delta\theta) &\approx \frac{1}{2} \delta\theta^2 |n_{41}|^2 |S_K (\omega_{14})|^2 \propto \delta\theta^2.
    \label{eq:L4}
\end{align}
\end{subequations}
A different method is required when considering the near-resonant leakage state $\ket{2}$. While the $\ket{1} \rightarrow \ket{2}$ transition is directly coupled by $\hat{n}$, with $|\matrixel{2}{n}{1}| \approx n_0 \sqrt2$, this leading-order term vanishes when our subsequence is applied, as demonstrated in Eq.~\eqref{eq:phase_cancellation}. At $n_g=  0$, even powers of $\delta \theta$ vanish by parity from the $\ket{1} \rightarrow \ket{2}$ amplitude. After cancellation of its linear term, that amplitude starts at cubic order, whereas the $\ket{0}\rightarrow\ket{2}$ amplitude starts at quadratic order. As such the single-period population of $\ket{2}$ at $n_g=0$ is suppressed to quartic order, $L_2 \propto \delta \theta ^4$. More specifically,
\begin{equation}
    L_2 \simeq \frac{\delta \theta ^4}{8} |n_{01} n_{12}|^2 |S_K (\omega_{02})|^2 \cot^2 \left( \frac{\omega_{01} \tau_K}{2} \right).
    \label{eq:L2}
\end{equation}
The coefficient above retains the dominant $\ket{1}$ intermediate state in the second-order $\ket{0}\rightarrow\ket{2}$ amplitude, and demonstrates how our technique achieves such strong suppression of leakage to $\ket{2}$ relative to leakage into $\ket{3}$ and $\ket{4}$. At fixed $\delta \theta$, $L_2$ nevertheless grows with $K$ for $K \geq 4$, by a factor of 12 from $K = 4$ to $K = 8$, because both $|S_K(\omega_{02})|$ and $\cot(\omega_{01}\tau_K/2)$ increase with $K$; this is the rise visible in Fig.~\ref{fig:leakage_K8}(b).

If $\tau_K$ = $T_{12}/K$, $S_K( \omega)$ vanishes whenever $f / f_{12}$ is an integer not divisible by $K$. When we set $f = f_{12}$ to achieve cancellation in the $\ket{1}, \ket{2}$ space, we also end up with additional cancellations at higher harmonics. More specifically, the $\ket{0} \rightarrow \ket{3}$ transition lies very close to one of these, as in the Duffing approximation
\begin{equation}
    E_n - E_0 \simeq n \hbar \omega_p - \frac{E_C}{2} n (n+1),
\end{equation}
and so $f_{03} \approx 3 f_{12}$. Thus, for values of $K$ that do not divide 3, the subsequence also cancels the $\ket{0} \rightarrow \ket{3}$ amplitude at first order. This is an approximation, but for our chosen parameters it is very close, and we have $f_{03}/f_{12} \approx 2.995$.

Resulting from this, the leading-order quadratic contribution to $L_3$ ends up with an unusually small prefactor. At larger $\delta \theta$, quartic contributions will eventually dominate, since the transition $\ket{1} \rightarrow \ket{3}$ is primarily allowed through processes such as $\ket {1} \rightarrow \ket{2} \rightarrow \ket{3}$, and so its squared magnitude contributes at fourth order. This explains the ``hitch'' in the $L_3$ curve as a crossover from quadratic to nearly quartic scaling. This notably does not apply to $L_4$, since there is no comparable interference condition suppressing the higher-level leakage channels.

\subsection{Numerical verification}
\label{sec:numerics}

To confirm these perturbatively-derived scaling laws, we can compute the leakage into each state numerically, using the period operator $\hat{U}_{\mathrm{per}}$. We construct the operators $\hat{U}_{\mathrm{free}} (t) = e^{-i \hat{H} t / \hbar}$ and $\hat{U}_{\mathrm{SFQ}}$ in the eigenstate basis of the transmon Hamiltonian, and we truncate the eigenbasis to, at a minimum, ten energy levels. The charge matrix elements $n_{kl}$ used for constructing $\hat{U}_{\mathrm{SFQ}}$ are obtained by numerically diagonalizing the transmon Hamiltonian in the charge basis with $|n| \leq N_{\mathrm{charge}} = 110$.

Except when stated otherwise, we use $E_C / h = 0.2 \text{ GHz}$ and $E_J / h = 15 \text{ GHz}$, and so $E_J / E_C = 75$. With this transmon, $\omega_{01} / 2 \pi = 4.690 \text{ GHz, } \omega_{12} / 2 \pi = 4.467 \text{ GHz }$ and $T_{01} = 0.213 \text{ ns, } T_{12} = 0.224 \text{ ns}$. The reason for operating so deep in the transmon limit is discussed in Sec.~\ref{sec:detuning}. Figure~\ref{fig:leakage_K8} uses a $K = 4$ subsequence to demonstrate a clear hierarchy of scaling laws: $\ket{2}$ is suppressed to quartic order for $K\geq2$ at $n_g=0$, whereas $\ket{3}, \ket{4}, \ket{5}$ scale with $\delta \theta ^2$ for small $\delta \theta$, as predicted by Eqs.~\eqref{eq:L3andL4} and \eqref{eq:L2}. The dominant leakage channel is $\ket{4}$, and the $L_3$ leakage crosses over into quartic scaling with $\delta \theta \gtrsim 2 \times 10^{-3}$, because $S_K(\omega_{03})$ is near zero and this strongly suppresses the leading-order quadratic term.

Figure~\ref{fig:leakage_K8}(b) shows the dependence on $K$ of leakage to each state. For a fixed value of $\delta \theta = 10^{-3}$, $L_3$ is of order $10^{-13}$ for $K = 2, 4, 6, 8$ but rises to approximately $5.4\times10^{-9}$ at $K = 3$. $L_4$ varies little with $K$ except for the order-of-magnitude spike at $K = 3$, because $|S_3 ( \omega_{14})| \approx 2.9$. When geometric suppression of $L_2$ becomes possible at $K = 2$, its value drops by eight orders of magnitude. Numerical resolution is sufficient even at the extremely small values towards the bottom of the figure: the plotted leakage amplitudes remain well above floating-point error. With quartic suppression in $\delta \theta$ of the leakage to $\ket{2}$ established alongside the quadratic scaling of the off-resonant leakage populations, we can now turn to demonstrating the gate times and fidelities resultant from application of our technique.

\section{Analysis of results}
\label{sec:results}

\subsection{Simulation details, fidelity definition, and calibration}
\label{sec:calibration}

We simulate the system dynamics by iterating the period operator $\hat{U}_{\mathrm{per}}$ [Eq.~\eqref{eq:u_period}], constructed as per Sec.~\ref{sec:numerics}, and recording the population of each energy eigenstate. The gate performance is given in terms of average gate fidelity $F_{\mathrm{avg}}$ in the computational subspace~\cite{Pedersen_2007}:
\begin{equation}
    F_{\text{avg}} = \frac{|\text{Tr}(\hat{U}_{\text{ideal}}^{\dagger} \hat{U}_{\text{actual}})|^2 + \text{Tr}(\hat{U}_{\text{actual}} \hat{U}_{\text{actual}}^{\dagger})}{d(d+1)},
    \label{eq:favg}
\end{equation}
where $\hat{U}_{\mathrm{ideal}}$ is the intended Pauli $X$ gate and $\hat{U}_{\mathrm{actual}}$ is the total evolution operator truncated to the computational subspace and rotated to the correct azimuth on the Bloch sphere, which corresponds to the previously-described virtual $Z$ gate~\cite{McKay_2017}. For the $X/2$ results, $\hat{U}_{\text{ideal}}$ in Eq.~\eqref{eq:favg} is instead $R_x(\pi/2)$ and optimized independent virtual-$Z$ rotations are added before as well as after the pulse sequence. The leakage error is taken into account in this definition of gate fidelity because $\mathrm{Tr}(\hat{U}_{\mathrm{actual}} \hat{U}^\dagger _{\mathrm{actual}} ) = d = 2$ only when $\hat{U}_{\mathrm{actual}}$ is unitary, which will not be the case when there is a nonzero population of non-computational states. Furthermore, we can characterize the leakage $\mathcal{L}$ by computing the total population loss from the computational subspace~\cite{Wood_2018}:
\begin{equation}
    \mathcal{L} = 1 - \frac{1}{2} \mathrm{Tr} \left( \hat{U}_{2 \times 2 } \hat{U}^\dagger_{2 \times 2} \right) = \sum_{n \geq 2} L_{n}.
    \label{eq:leakage_metric}
\end{equation}
Here $L_n$ is computed over the entire gate. For a fixed $K$ and $\delta \theta$ the sequence is calibrated by determining the number of periods $N$ required to complete the rotation and determining the compensatory period detuning $\alpha$ from Eq.~\eqref{eq:trep}. We search integer values of $N$ near the first computational rotation and minimize $1-F_{\text{avg}}$ with respect to $\alpha$. The virtual-$Z$ correction is then optimized for each calibrated gate. The detuning is important: for $K = 4 \text{, } \delta \theta = 0.045$ the optimally detuned sequence ($\alpha = 1.85 \times 10^{-4}, N = 136, T_{\mathrm{gate}} = 29.0 \text{ ns}$) achieves an error of $4.6 \times 10^{-6}$, whereas the identical sequence with $\alpha = 0$ can only achieve an error of $1.7 \times 10^{-3}$: the off-resonance condition becomes significant as the phase mismatch accumulates over the $N$ periods. Appendix~\ref{app:alpha} shows the optimized values of $\alpha$ for $K \in \{1, 2, 4, 6, 8\}$. For each sampled $K$, at small kick strengths the $X$-gate detuning follows $\alpha = c_K \delta \theta^2 $ to a very close approximation. The coefficients $c_K$ are positive for all $K \neq 1$, and are listed in Appendix~\ref{app:alpha}. This quadratic dependence is to be expected from the second-order correction to the effective Floquet Hamiltonian generated by the kicks, also covered in more detail in Appendix~\ref{app:alpha}.

\subsection{Leakage and gate speedup}

\begin{figure}[htbp]
  \centering
  \includegraphics[width=\columnwidth]{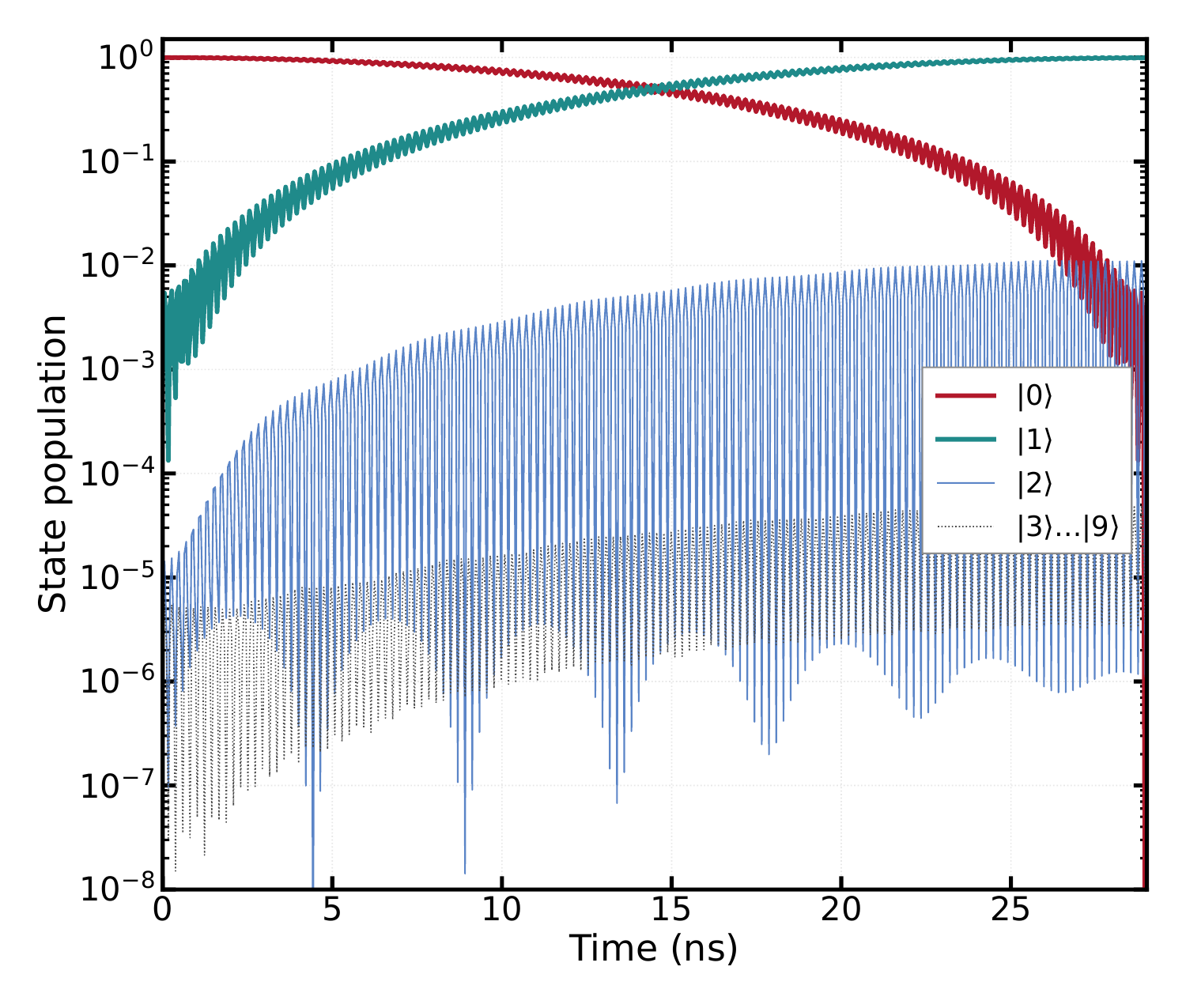}
  \caption{Time evolution of state populations throughout the gate as an $X$ rotation from $\ket{0}$ to $\ket{1}$ is effected. Here $E_C/h = 0.2$~GHz, $E_J/h = 15$~GHz, $n_g = 0$, $K = 4$ kicks per period, and $\delta\theta = 0.045$, with a gate time of 29 ns. There is a smooth, coherent Rabi-like transition from $\ket{0}$ to $\ket{1}$ exhibited in the computational subspace. $\ket{2}$, the first leakage state, exhibits high-frequency intra-period oscillations, demonstrative of the `balancing' of the geometric cancellation over one complete period. The population of $\ket{2}$ transiently reaches $(\delta \theta |n_{12}| / \sin(\pi /K))^2 P_1 \approx 1.1 \times 10^{-2} P_1$, but returns to values of $\sim 10^{-6}$ or smaller at period boundaries, with the envelope growing along with the population of $\ket{1}$. The gray dotted curve is the summed populations of the remaining noncomputational states included in the numerical calculations.}
  \label{fig:state_populations}
\end{figure}

It was demonstrated in Section~\ref{sec:single_period_leakage} that the geometric cancellation suppresses single-period leakage $L_2$ into state $\ket{2}$ at $n_g=0$ to quartic order in the SFQ strength $\delta \theta$. Figure~\ref{fig:state_populations} depicts the resulting dynamics via a representative example. We iterate through each operator, evaluating $P_f(t) =| \langle f | \psi(t) \rangle |^2$ for $\ket{f} \in \{ \ket{0}, \ket{1}, \dots, \ket{9} \}$ as we go. Here we use $K = 4, \, \delta \theta = 0.045$ and calibration gives $\alpha = 1.85 \times 10^{-4}, \, N = 136$. The gate duration is $29.0 \text{ ns}$. Within the computational subspace we observe the expected coherent population transfer from $\ket{0}$ to $\ket{1}$.  $P_2$ exhibits fast oscillations within each period, the upper envelope of which grows from $\sim 10^{-5}$ to $\sim 10^{-2}$ over the gate's evolution. The upper envelope of $P_2$ is \textit{not} accumulated leakage, as it returns to its minimum within the envelope after each period due to geometric cancellation. Specifically, the first-order amplitude of the $\ket{1} \rightarrow \ket{2}$ transition traces a closed, regular $K$-gon in the complex plane (Eq.~\eqref{eq:phase_cancellation}). Thus (for $K = 4$) the maximum excursion from the origin will be $\approx \sqrt{2} \delta \theta |n_{12}|$, corresponding to a maximum transient population of $\ket{2}$ of around $P_1(t) ( \delta\theta |n_{12}|/\sin(\pi/K))^2$, which for these parameters is $\sim 0.01 P_1(t)$. It can also be seen that this grows as the gate progresses due to the increase in $P_1 (t)$. The transient values are not a problem for gate fidelity without errors in the pulse train, since the transient population occurs \textit{during} a subsequence and an integer number of subsequences is always applied. A physical cost of this is the relaxation of $\ket{2}$ during its transient occupation~\cite{Abad2025DecoherenceLeakage}. With a time-averaged $\bar{P}_2 \approx 3 \times 10^{-3}$ and a relaxation time of around $T_1 / 2$ for state $\ket{2}$, the additional gate error is estimated to be of order $2 \bar{P}_2 T_{\mathrm{gate}} / T_1 \approx 2 \times 10^{-6}$ for a representative value of $T_1 = 100 \, \mu \mathrm{s}$. This is nearly two orders of magnitude below the decoherence floor given in Eq.~\eqref{eq:decoherence}. In Sec.~\ref{sec:jitter} and Appendix~\ref{app:pulse_drops}, we discuss in further detail the consequences of pulse train errors that leave this polygon unclosed, such as timing jitter or stochastic pulse drops, which leave a residual amplitude of $O( \delta \theta |n_{12} |)$ in $\ket{2}$.

\begin{figure*}[t]
    \centering
    \includegraphics[width=0.98\textwidth]{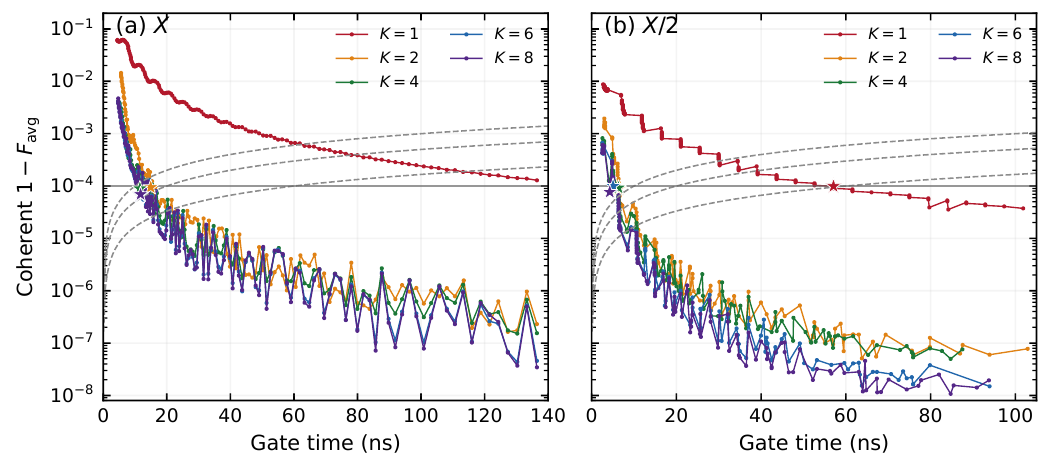}
    \caption{Coherent $1-F$ for (a) $X$ and (b) $X/2$ gates as a function of gate time, with $K = 1, 2, 4, 6 \text{ and } 8$. Both panels are calculated with the same 20-level transmon model. Each $K$ is sampled at 150 logarithmically spaced kick strengths, using the same grid for both panels. The subsequence count $N$, detuning $\alpha$ and virtual-$Z$ corrections are optimized for the target gates. Stars denote the fastest sampled sequences achieving $1 - F < 10^{-4}$. Gray dashed lines indicate first-order decoherence estimates for $T_1 = T_2 = 50, 100 \text{ and } 300 \, \mu\text{s}$. The horizontal line demarcates $1 - F = 10^{-4}$.}
    \label{fig:fidelity_vs_gatetime}
\end{figure*}

Figure~\ref{fig:fidelity_vs_gatetime}(a) summarizes the numerically derived gate performance resulting from our sequences. We plot the calibrated gate error $1 - F_{\mathrm{avg}}$ versus $T_{\mathrm{gate}}$ for $K \in \{ 1, 2, 4, 6, 8\}$. Each point corresponds to a sequence with a different value of $\delta \theta$, ranging from 0.002 to 0.065 for $K = 1$ and 0.005 to 0.41 for $K \geq 2$ (the large-$\delta \theta$ end, up to $C_C/C_\Sigma = 0.066$, is shown to expose the scaling rather than as a buildable device). $N$ and $\alpha$ are optimized for each sequence as described previously. Because $N \in \mathbb{N}$, the error spread of approximately one decade at the long-$T_{\mathrm{gate}}$ end of each curve is affected by what we term \textit{granularity}, the tendency of the sequence to overshoot or undershoot the desired $\pi$ rotation. We see that the $K = 1$ sequence -- the conventional SFQ drive -- is strongly limited by leakage as is expected since the $\ket{1} \rightarrow \ket{2}$ near-resonant drive is not canceled, and reaches $10^{-3}$ at approximately 50 ns but does not reach $10^{-4}$ within 130 ns. In contrast to the single-pulse drive, all sequences with $K \geq 2$ exhibit substantially better characteristics: the error first falls below $10^{-4}$ at approximately 11.5--14.9 ns in the sampled sequences; it first falls below $10^{-5}$ at 19--30 ns and remains below $10^{-5}$ for all sampled multi-kick $X$ gates with $T_{\mathrm{gate}}>43$ ns. It is notable that at a coherent error threshold of $10^{-4}$, increasing $K$ from 2 to 8 shortens the gate by only around $15 \text{ to }25 \%$, whereas the peak pulse rate that the control electronics must deliver increases from $\sim 10 \text{ to } 58 \text{ GHz}$, and the kick strength required for a given gate time scales inversely with $|S_K (\omega_{01})|$ since the kicks partially cancel on the $\ket{0} - \ket{1}$ circle due to the near-resonance characteristic of transmons. (To illustrate, for $K = 2, 4, 6, 8$, we have $|S_K(\omega_{01})| \approx 0.16, 0.21, 0.30, 0.39$.) Nonetheless, larger $K$ can lower leakage for equivalent gate time, since the per-kick error budget is smaller, and with increased $K$, we suppress additional harmonics of the $\ket{1} \rightarrow \ket{2}$ transition. In Sec.~\ref{sec:resonances} we discuss how this also decides the offset-charge sensitivity.

To leading order in $\delta \theta$, the transverse computational rotation angle per subsequence is

\begin{equation}
    \Theta \simeq 2 \delta \theta |n_{01}| |S_K (\omega_{01} ) |.
    \label{eq:bigtheta}
\end{equation}
An $X$ gate requires $N \Theta \approx \pi$. Using Eq.~\eqref{eq:dtheta}, we obtain

\begin{equation}
    T_{\mathrm{gate}} \simeq \frac{T_{\mathrm{rep}}}{4 |n_{01}| |S_K(\omega_{01})|} \frac{C_\Sigma}{C_C}.
    \label{eq:gatetime_capacitance}
\end{equation}
As such, while the coupling capacitance required for a given gate time depends on $K$, the geometric cancellation condition does not. The value of $K$ can therefore be altered post-fabrication without altering $C_C$, although this will produce a different gate time and require recalibration of $N$ and $\alpha$.

There is more freedom in setting the repetition-period detuning $\alpha$ with the $X/2$ gate than with the $X$ gate. An $X$ gate requires a rotation axis in the equatorial plane, whereas an $X/2$ gate can be implemented using a rotation about a tilted axis while applying virtual-$Z$ rotations before and after the sequence. The $X/2$ sequences shown in Fig.~\ref{fig:fidelity_vs_gatetime}(b) first reach coherent error values below $10^{-4}$ at $T_{\text{gate}} \approx 4.3 \text{ to } 6.3 \text{ ns}$ with $K \geq 2$, compared with approximately $11.5 \text{ to } 14.9 \text{ ns}$ for $X$ gates. This is described in further detail in Appendix~\ref{app:xhalf}.

In subsequent examples, we use $K = 4$ as a compromise, giving a peak pulse rate of $22 \text{ GHz}$ (and thus minimal pulse spacing of 45 ps, accounting for $\tau_{\text{last}}$ as given in Eq.~\eqref{eq:tau_last}). This lowers the required peak pulse rate by a factor of 2.6 relative to $K = 8$~\cite{Jordan2026} but also restricts the minimal value that $E_J/E_C$ can take. We achieve a gate time of around 30 ns and coherent error that is significantly below the decoherence floor. We could have chosen $K = 3$, and may well have for other qubit parameters, but this is strongly disfavored by the constructive interference in the $\ket{0} \rightarrow \ket{3}$ transition for our specific qubit.

We can now discuss the \textit{incoherent} error accumulated over the gate. We have a long relaxation time $T_1$ and dephasing time $T_2$ compared to $T_{\text{gate}}$, where $1 / T_2 = 1 / (2 T_1) + 1 / T_{\varphi}$. As such we can calculate to first order the contribution to the gate $1 - F$ from incoherent error by the following method. We apply over the gate time Eq.~\eqref{eq:favg} to sum over the Kraus operators describing amplitude damping ($T_1$) and pure dephasing ($T_\varphi$) \cite{Gambetta_2011, Korotkov_2013, OMalley_2015} and determine the contribution to the average gate infidelity of
\begin{equation}
    1 - F_{\text{dec}} \simeq \frac{T_{\text{gate}}}{3} \left( \frac{1}{2 T_1} + \frac{1}{T_2} \right),
    \label{eq:decoherence}
\end{equation}
which is approximately equal to $T_{\text{gate}} /( 2 T_1 )$ when $T_2 \approx T_1$. This decoherence floor is shown in Fig.~\ref{fig:fidelity_vs_gatetime} for $T_1 = T_2 = \text{50, 100 and 300 } \mu\text{s}$ (that is, $T_\varphi = 2 T_1$) as dashed lines. For $X$ gates with $K \geq 2$, the coherent and incoherent errors cross around 15 ns. Minimizing their sum at $T_1 = T_2 = 100 \, \mu\text{s}$ gives a total value of $1 - F$ between $1.0 \times 10^{-4}$ and $1.3 \times 10^{-4}$, with gate times of 16 to 21 ns. With the decidedly suboptimal $K = 1$ gate, we have a best $1-F$ of $7.4 \times 10^{-4}$ at approximately 95 ns in a denser sweep. Thus for $K\geq2$, decoherence dominates after crossover, and as such the much lower coherent errors at longer gate times are inaccessible. Nonetheless, with higher $K$, the gate can be short enough for the decoherence floor to be the primary source of error, and leakage can thus be rendered negligible.

In short, even with decoherence, our protocol allows fast $X$ gates (16 to 21 ns) with error $1 - F$ values in the vicinity of $10^{-4}$ without introducing drastic hardware complication. Whether this clears a given fault-tolerance threshold depends on the full error model, to which we return in Sec.~\ref{sec:conclusion}. The remainder of the paper will discuss the error channels limiting the performance of the calibrated gate: detuning of the drive frequency from the dressed frequency due to offset charge or clock drift (Sec.~\ref{sec:charge_noise}) and timing jitter in the SFQ pulse train (Sec.~\ref{sec:jitter}).

\section{Frequency mismatch and resonant leakage}

\label{sec:charge_noise}

The sequence is initially calibrated to a given value of the dressed qubit frequency, which can change over the qubit's lifespan. Changes in qubit characteristics have two effects: the first and most obvious is the detuning of the qubit frequency $f_{01}$, which produces a simple rotation error that is smooth in the drift, accurately described by a single parameter, and is discussed in Sec.~\ref{sec:detuning}. The second is changes in the higher-level transition frequencies, which can bring transitions out of the computational subspace into resonance with harmonics of the erroneous pulse train. This leads to very high but narrow leakage peaks at particular offset charges, and we discuss it in Sec.~\ref{sec:resonances}. Finally, we discuss the resolution, or granularity, of the calibration itself in Sec.~\ref{sec:granularity}.

\subsection{Qubit frequency mismatch}
\label{sec:detuning}

When implementing our method on hardware, the control sequence is calibrated once at a known offset charge, which fixes the qubit frequency to which we match the clock. The pattern is then held fixed through the gate, with $\delta \theta$ set at fabrication (Eq.~\eqref{eq:dtheta}) and the pulse spacing $\tau_K$ and subsequence repetition period $T_{\text{rep}}$ set by the clock at calibration. Drift of the clock itself, the offset charge, or the qubit frequency (due to two-level defects and/or flux noise) creates a mismatch $\epsilon$ in the sequence:
\begin{equation}
    \epsilon =
    \frac{[\omega_{01}T_{\mathrm{rep}}]_{\text{op}}}{2 \pi (1+\alpha)} - 1 = \frac{[\omega_{01}T_{\mathrm{rep}}]_{\text{op}}}{\omega_{01} T_{\text{rep}} } - 1,
    \label{eq:eps}
\end{equation}
where the subscript $\mathrm{op}$ denotes quantities evaluated as the qubit is operated, $[\omega_{01}] _{\text{op}}$ is the bare qubit transition frequency, and $\alpha$ is held at its calibrated value. This normalization gives zero mismatch at calibration. For the offset charge case, the charge dispersion of $f_{01}$ scales exponentially with $\sqrt{8 E_J / E_C}$~\cite{Koch_2007,Serniak2019ChargeParity}:
\begin{equation}
    \Delta f_{01} \propto \exp \left( - \sqrt{8 E_J / E_C} \right).
    \label{eq:f01}
\end{equation}
At a fixed value of $E_J / h = 15 \text{ GHz}$ we have a peak-to-peak dispersion of $f_{01}$ over $n_g$ equal to 0.15, 1.9 and 12 kHz at $E_C / h =$ 0.20, 0.25 and 0.30 GHz respectively; mismatches of $\epsilon = 3.2 \times 10^{-8}, 3.6 \times 10^{-7} \text{ and } 2.1 \times 10^{-6}$. Since the gate consists of $N \approx 100$ periods, even a small value of $\epsilon$ results in a $2 \pi N \epsilon$ phase between the drive and the qubit over the entire gate operation.

\begin{figure}[t]
    \centering
    \includegraphics[width=\columnwidth]{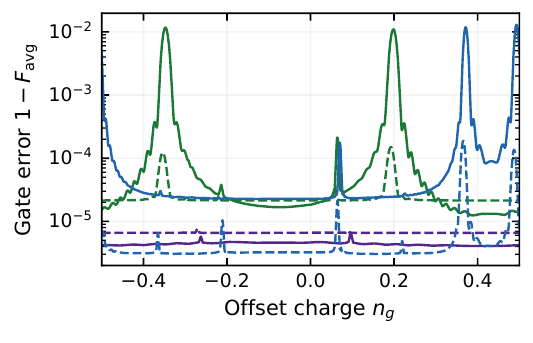}
    \caption{Gate error as a function of the static offset charge $n_g$. Here $E_J / h = 15$ GHz, and we have $E_J / E_C = 75$ (purple), 60 (dark green) and 50 (blue). Solid lines are for $K = 4$, dashed lines are for $K = 8$. Gate times are approximately 29, 24 and 21 ns, respectively, for the three values of $E_C/h$, regardless of $K$. The sharp peaks are due to resonant leakage into $\ket{6}$ via harmonics of the entire sequence, which is substantially suppressed by increasing $K$ to 8. Clock timings and virtual-$Z$ corrections are held fixed during each scan at the values calibrated at $n_g=0$.}
    \label{fig:fidelity_vs_ng}
\end{figure}

These sequences are calibrated for $n_g = 0$. Figure~\ref{fig:fidelity_vs_ng} depicts the effect of a nonzero $n_g$ at $E_J / E_C = $ 75, 60 and 50. The kick strengths, $\delta \theta = 0.045$ for $K = 4$ and 0.0245 for $K = 8$, give 29-ns gates at $E_C/h = 0.20$ GHz and are held fixed, so the gates are shorter at the larger $E_C$ (see the caption). Away from the higher-level resonances discussed in the following subsection, the error is typically of order $10^{-5}$ or smaller. For small frequency mismatches ($|\epsilon|N\ll 1$), we have

\begin{equation}
    1 - F_{\text{avg}} \simeq (1 - F_0) + \frac{8}{3}\, (\epsilon N)^2,
    \label{eq:detuning}
\end{equation}
where $F_0$ is the calibrated fidelity, and everything else is as given before. Due to the small values of $\epsilon$ shown above, the quadratic dependence means that the effect on the gate error is negligible. At $E_C/h = 0.20 \text{ GHz}$ we would need a charge dispersion greater by a factor of 300 to double the gate error, and at 0.30 GHz, the worst case, it would still need to be greater by a factor of 10 (for that gate, $N = 118$ and $1 - F_0 = 2.3 \times 10^{-5}$). As such, the offset charge drift itself is not an issue except where it intersects with the error peaks created by unintended higher-level resonances.

\begin{figure}[t]
    \centering
    \includegraphics[width=\columnwidth]{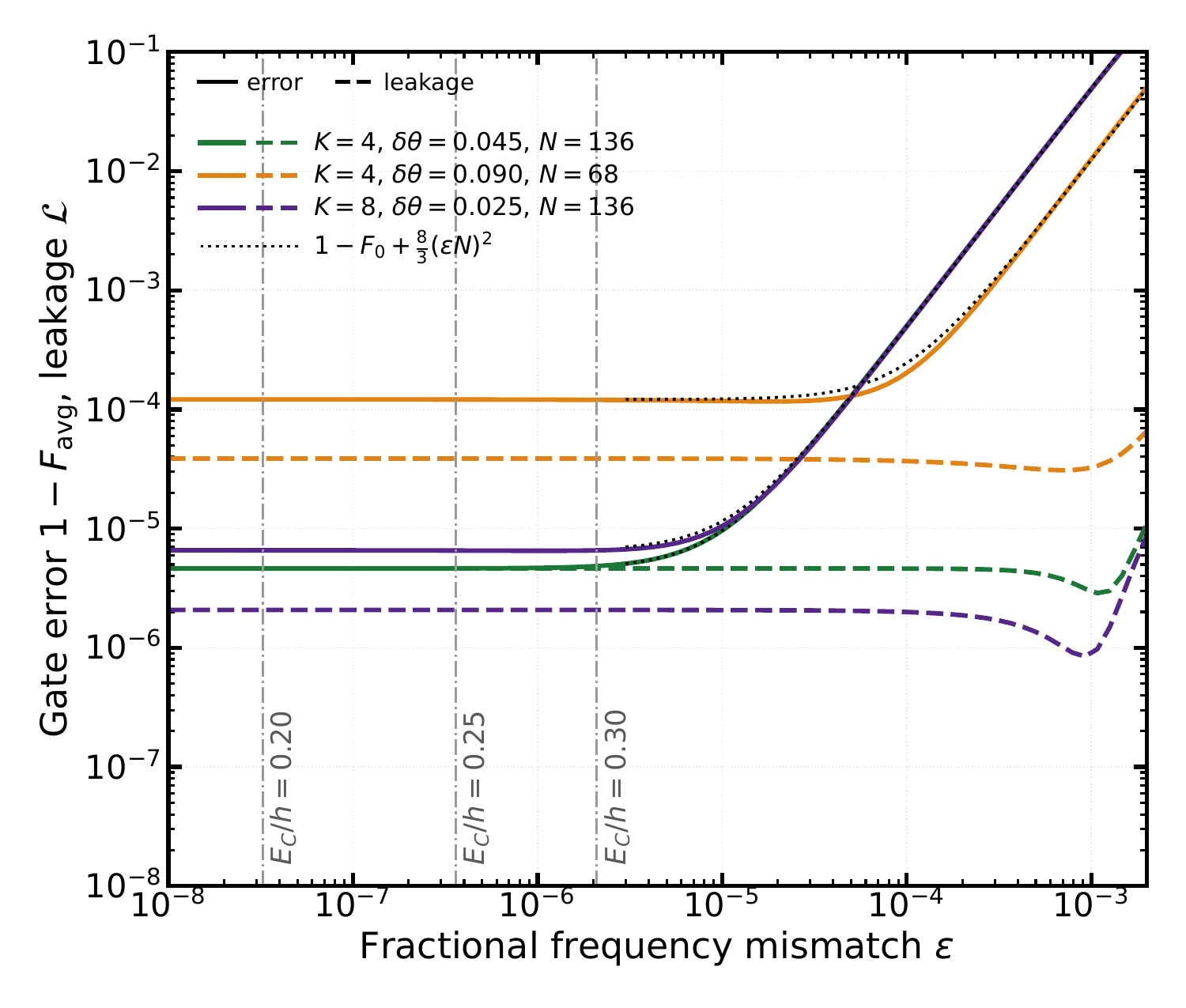}
    \caption{Gate error $1 - F$ (solid lines) and leakage $\mathcal{L}$ (dashed lines) shown versus the fractional clock mismatch $\epsilon$. The sequences are $K = 4, N = 136; \, K = 4, N = 68 \text{ with a higher value of } \delta \theta \text{, and } K = 8, N = 136$. These sequences achieve gate times of 29 ns, 14.5 ns and 29 ns respectively. The dotted lines are calculated from Eq.~\eqref{eq:detuning}. Vertical lines indicate peak-to-peak computational charge dispersion for $E_J / E_C = 75, \, 60 \text{ and } 50$.}
    \label{fig:detuning}
\end{figure}

Figure~\ref{fig:detuning} shows the leakage and total error of three calibrated gates as a function of the clock detuning $\epsilon$. We leave the virtual-$Z$ angle $\phi_z$ unchanged from the original calibrated value, which contributes non-leakage error, and the gate parameters are detailed in the figure caption. For small mismatches the error is described accurately by Eq.~\eqref{eq:detuning}. This detuned drive causes the state to be rotated about an unintended axis, outside of the equatorial plane; the angle of this tilted axis is $\beta = \arctan 2\epsilon N$ and a $\pi$ rotation about it introduces an error cost of $(1 - F)_{\beta} = ( 2/3) \sin^2 \beta$. Our error budget for a given maximal permissible error $E_{\text{max}}$ is then
\begin{equation}
    |\epsilon| < \frac{1}{N} \sqrt{\frac{3}{8} \left( E_{\text{max}} - (1 - F_0) \right)}.
    \label{eq:epsilon_condition}
\end{equation}
For the 29-ns gates we thus need $|\epsilon|\lesssim4.4\times10^{-5}$ for $E_{\max}=10^{-4}$. For the 14.5-ns gate, $|\epsilon| < 8.7\times10^{-5}$ instead corresponds to an \emph{additional detuning-induced error} budget of order $10^{-4}$: its zero-mismatch infidelity is already approximately $1.22\times10^{-4}$. The offset charge dispersion of $f_{01}$ with $E_C / h = 0.2 \text{ GHz}$ is three orders of magnitude below these values $(|\epsilon| = 3.2 \times 10^{-8})$, and so the frequency stability of the system is found to set the practical error limit.

The leakage $\mathcal{L}$, unlike the total error, remains within a factor-of-two band up to the high $|\epsilon|$ value of around $10^{-3}$ (where total error has grown by four orders of magnitude). A rescaling of \textit{all} times preserves the equal-spacing pattern but detunes the geometric cancellation of the SFQ kicks and the residual comb factor $|S_K (\omega_{12})| \simeq \pi |\epsilon| / \sin( \pi / K)$ then grows only linearly with $\epsilon$. The protocol thus retains the intended leakage suppression, fixed by the pulse pattern, while error within the computational subspace is determined by the phase coherence between the qubit and the SFQ clock. The import of targeting leakage specifically arises from the fact that when performing actual error-corrected multiple-qubit computations, standard stabilizer codes do not in general detect or correct leakage~\cite{Krinner2022SurfaceCode,Marques2023LeakageReduction,Lacroix2025LeakageReduction}. An increase in error from qubit frequency drift can thus be compensated in a far easier manner because the leakage suppression is not substantially impeded.

\subsection{Higher-level resonances and leakage}
\label{sec:resonances}

In those unfortunate cases where higher harmonics of the period-shifted pulse sequence are in resonance with higher-level transitions, i.e., $\ket{1} \rightarrow \ket{6}$, $\ket{1} \rightarrow \ket{4}$ or $\ket{0} \rightarrow \ket{3}$, we have sharp peaks in leakage. Fig.~\ref{fig:fidelity_vs_ng} demonstrates these, which rise to nearly three orders of magnitude above the calibrated error. Unlike the intra-period resonance peaks demonstrated in Sec.~\ref{sec:single_period_leakage}, which apply over a subsequence, these build up over the entire sequence. These exist because the higher levels move with the offset charge: between $n_g = 0$ and $n_g = 0.5$, the transition frequency $f_{16}$ shifts by 0.21, 0.84 and 1.81 GHz at $E_C / h = $ 0.20, 0.25 and 0.30 GHz respectively. At $n_g=0$, the energy gaps between $\ket{6}$ and the top of the barrier at these three charging energies are 3.5, 1.7 and 0.3 GHz respectively. The deep levels are exponentially insensitive to $n_g$, whereas the shallow ones are not.

We can demonstrate that for the $\ket{1} \rightarrow \ket{n}$ transition over an entire gate, the accumulated amplitude can, to first order, be factorized as
\begin{equation}
    |A_{1n}^{(1)}| = \delta\theta\, |n_{1n}|\, \bigl|S_K(\omega_{1n})\bigr| \, D_N(\omega_{1n}),
    \label{eq:rescond_a}
\end{equation}
where $S_K$ [Eq.~\eqref{eq:comb}] is the interference of the $K$ kicks within one period and $D_N$ is that of the $N$ periods of the gate,
\begin{equation}
    D_N(\omega) = \left|\frac{\sin(N \omega T_{\text{rep}} /2)}{\sin(\omega T_{\text{rep}}/2)}\right|,
    \label{eq:rescond_b}
\end{equation}
and beyond strict first order we approximately replace the argument of $D_N$ by the dressed frequency $\tilde{\omega}$. We use tildes to denote such dressed frequencies throughout, and discuss the origin of these frequency shifts further in Appendices~\ref{app:alpha} and \ref{app:res16}. Note that Eq.~\eqref{eq:rescond_b} is identically the equation for the interference factor of a grating of slits in an optical context. This factorization ignores the changing population of $\ket{0}$ and $\ket{1}$ as the gate progresses, but does capture the interference conditions that are responsible for the higher-level resonances. The effect of the rotation in the computational states on these amplitudes is treated in Appendix~\ref{app:res16}.

$S_K$ is the interference of the $K$ kicks of one period, and vanishes only when $f_{1n}T_{12}$ is an integer that $K$ does not divide. It sets the size of the amplitude per period. $D_N$ is the interference of all $N$ periods and is very sharp: its maxima are reached when $\tilde{f}_{1n}T_{\text{rep}}$ is an integer. For this interference factor, the distance from each maximum to its first zero is $1/N$ in $\tilde{f}_{1n}T_{\text{rep}}$, i.e., $1/(NT_{\text{rep}}) \approx 34$ MHz in $\tilde{f}_{1n}$ for the gate of Fig.~\ref{fig:state_populations}. Significant resonant leakage requires $\tilde f_{1n}T_{\mathrm{rep}}$ near an integer and $|S_K(\omega_{1n})|$ appreciably nonzero; $S_K$ need not lie at a principal maximum. The condition $f_{1n}T_{12}\approx\ell K$ identifies a possible principal maximum, not a necessary resonance condition.

For the sequences of Fig.~\ref{fig:fidelity_vs_ng}, the $\ket{1} \rightarrow \ket{6}$ transition has both $S_K$ and $D_N$ take large values, as the effective dressed transition frequency is near the fourth harmonic of the repetition frequency ($\tilde{\omega}_{16} T_{\text{rep}} / 2 \pi \approx 4$). For $K = 4$, we also have $|S_4(\omega_{16})| \approx 3.6$, which is to say that the fourth harmonic of the repetition frequency is close to a maximum of the \textit{intra}-period comb. The resulting transition amplitude per period is then approximately $\delta \theta | n_{16} | |S_4 (\omega_{16} )| \approx 1.3 \times 10^{-3}$ -- not particularly large only because the charge matrix element is small, specifically $| n_{16} | \approx 0.008$. The amplitudes contributed by successive periods add coherently, producing a leakage population in $\ket{6}$, the growth of which is weighted by the evolving amplitude of $\ket{1}$. Assuming a $\pi$ rotation in the computational subspace and coherent accumulation in $\ket{6}$, the approximate peak population of $\ket{6}$ can then be shown to be
\begin{equation}
    P_{6, \text{ peak}} \approx \left( \frac{2N}{\pi} \delta \theta |n_{16} | |S_K ( \omega_{16} )| \right)^2.
    \label{eq:peak_higher_leakage_6}
\end{equation}
For the 24 ns gate with $K = 4$, $E_C / h = 0.25$ GHz, Eq.~\eqref{eq:peak_higher_leakage_6} predicts $P_6 \approx 1.0 \times 10^{-2}$, which agrees with our numerically obtained value of $1.1 \times 10^{-2}$ (gate error $1.2 \times 10^{-2}$). The subsidiary peaks that can be seen in Fig.~\ref{fig:fidelity_vs_ng} are the additional interference fringes associated with the full $N$-period sequence, and their characteristic frequency spacing is thus approximately $1/T_{\text{gate}}$. At $E_C/h = 0.20$ GHz, by contrast, the bare product $f_{16}T_{\text{rep}}$ stays between 4.14 and 4.19 over the whole period of $n_g$ and never reaches 4, which explains that curve's lack of a large peak.

It also makes sense to discuss the displacement of the resonance pattern due to asymmetry in the offset charge induced by the SFQ drive. The time-averaged gate charge contribution from the drive can be represented as
\begin{equation}
    \delta n_g = \frac{\hbar}{8 E_C}\frac{K \delta \theta}{T_{\text{rep}}}.
    \label{eq:delta_n_g}
\end{equation}
For the sequences we are using $\delta n_g$ generally takes values around 0.07-0.09, which is consistent with the displacement of the numerically calculated resonance patterns.

We can suppress these higher-level resonances by increasing the subsequence pulse count $K$. For each odd principal maximum of a $K$-kick comb, a $2K$-kick comb replaces it with a zero: $S_{2K}(\omega) = (1 + e^{i \omega \tau_K / 2}) S_K (\omega)$. In the case discussed above, going from $K = 4$ to $K = 8$ suppresses the resonance that leads to an abnormally high population of $\ket{6}$: although the cancellation is not exact for this qubit, the comb factor at the relevant resonance peaks is reduced from around 3.6 to around 0.6, each evaluated at the offset charge of its own peak. This in turn reduces the error (when $E_C/h = 0.25 \text{ GHz}$) from around $1.2 \times 10^{-2}$ to $1.5 \times 10^{-4}$. Table~\ref{tab:Kscan} collects the worst-case error over the full period of $n_g$ for 29-ns gates at each $K$ and $E_C$. Unlike in Fig.~\ref{fig:fidelity_vs_ng}, here we adjust the kick strength as we vary $E_C$ to maintain an approximately constant gate time. The $K = 8$ sequence has its first positive-frequency principal maximum at $8 \omega_{12}$. Increasing the truncation from 10 to 20 levels changes the sampled worst-case errors in Table~\ref{tab:Kscan} by less than 1\%. In general, since the higher-level resonances are strongly dependent on the transmon parameters, a programmable value of $K$ can allow for suppression of these idiosyncratic resonances without having to throw out qubits.

\begin{table}[htbp]
\caption{Worst-case gate error over a full period of the offset charge, $\max_{n_g \in [-1/2, 1/2]} (1 - F_{\text{avg}})$, of $\approx 29$-ns gates calibrated at $n_g = 0$, for $E_J/h = 15$~GHz and different numbers of kicks per period $K$. For each combination of $K$ and $E_C$, the kick strength is chosen to give approximately the same gate time; the dominant leakage channel at the worst $n_g$ is $\ket{4}$ for $K = 3$ at $E_C/h = 0.20$~GHz and $\ket{6}$ in all other cases where the error exceeds $10^{-4}$. The maximum is evaluated over 1001 uniformly spaced offset charges, at fixed pulse timings and virtual-$Z$ correction.}
\label{tab:Kscan}
\footnotesize
\begin{ruledtabular}
\begin{tabular}{cccc}
 & \multicolumn{3}{c}{$E_C/h$ (GHz), $E_J/E_C$} \\
$K$ & 0.20, 75 & 0.25, 60 & 0.30, 50 \\
\hline
3 & $1.5 \times 10^{-4}$ & $1.0 \times 10^{-3}$ & $1.4 \times 10^{-3}$ \\
4 & $1.6 \times 10^{-5}$ & $1.2 \times 10^{-2}$ & $1.3 \times 10^{-2}$ \\
5 & $1.7 \times 10^{-5}$ & $1.3 \times 10^{-3}$ & $1.9 \times 10^{-3}$ \\
6 & $3.0 \times 10^{-5}$ & $3.3 \times 10^{-4}$ & $4.7 \times 10^{-4}$ \\
8 & $7.9 \times 10^{-6}$ & $1.3 \times 10^{-4}$ & $1.8 \times 10^{-4}$ \\
\end{tabular}
\end{ruledtabular}
\end{table}

\subsection{Granularity of the calibrated gate}
\label{sec:granularity}

\begin{figure}[t]
    \centering
    \includegraphics[width=\columnwidth]{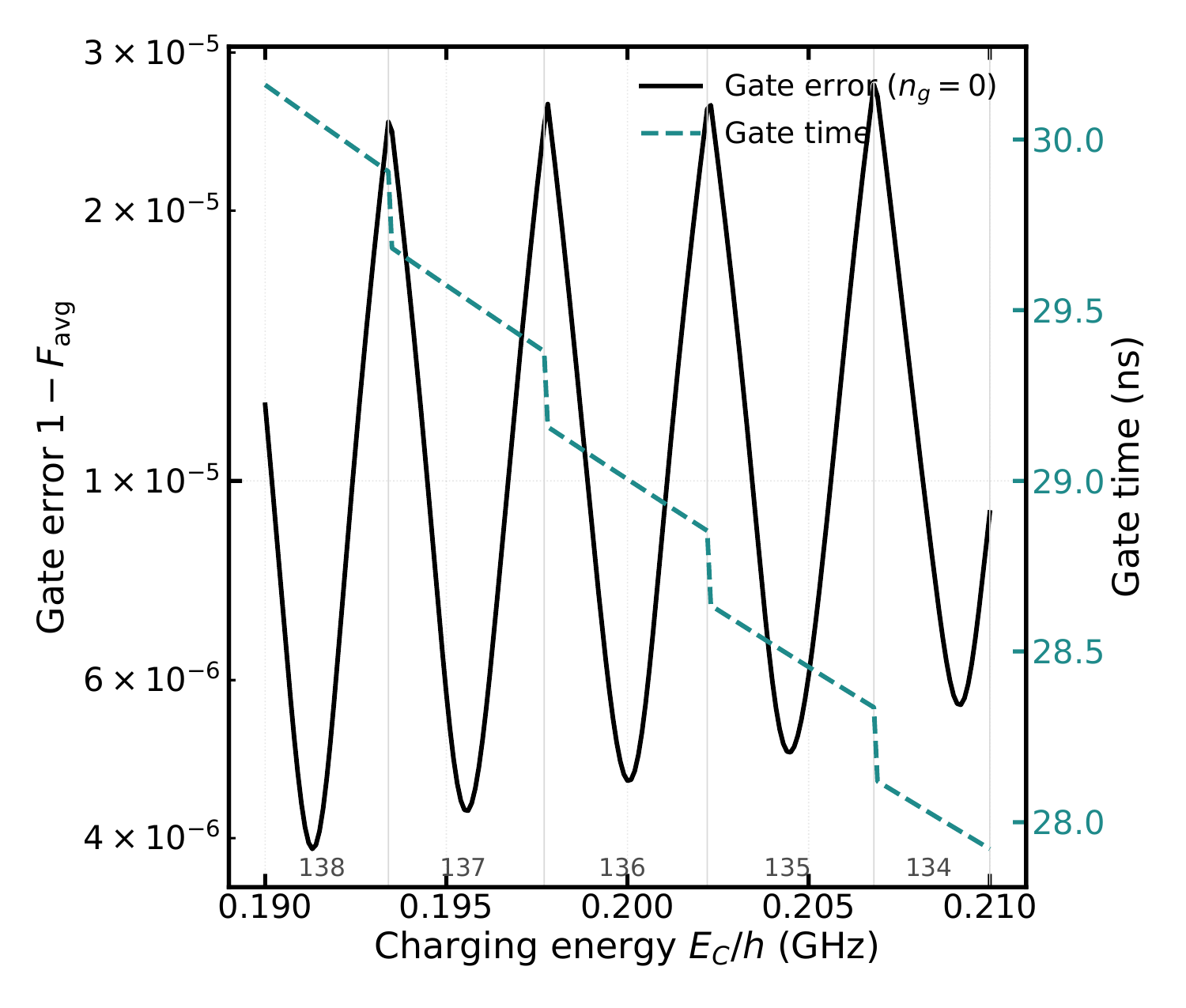}
    \caption{Calibrated gate error (left axis) and gate time (right axis) as functions of the charging energy $E_C$ for $K = 4$, $\delta \theta = 0.045$ and $n_g = 0$. The optimal value of $N$ changes discretely, which produces jumps in the gate time and oscillations in the gate error due to granularity, as described in Sec.~\ref{sec:granularity}.}
    \label{fig:granularity_vs_EC}
\end{figure}

The total gate is always effected over an integer number of periods, which implies that the sequence is comprised of an integer number of subsequences. This introduces a granularity error into the eventual gate, defined as the undershoot or overshoot of the final statevector from the targeted latitude on the Bloch sphere. Figure~\ref{fig:granularity_vs_EC} shows this effect via varying $E_C$ while holding all other relevant parameters ($E_J$, $\delta \theta$, $n_g$ and so on) fixed. As $E_C$ increases, the qubit period $T_{01}$ decreases, which speeds up the gate. However, the optimal value of $N$ changes discretely, introducing steps in $T_{\text{gate}}$.

If we state that each subsequence produces an approximate computational rotation $\Theta$, the period count $N$ that brings $N \Theta \approx \pi$ (for our $\pi \text{-gate}$) will leave an overshoot or undershoot of magnitude no greater than $\Theta /2$. For small rotation errors, we can state an approximate bound for the resulting gate infidelity from granularity as
\begin{equation}
    (1-F)_{\text{gran}} \lesssim \frac{1}{6} \left( \frac{\Theta}{2} \right)^2 = \frac{\Theta ^2}{24} \approx \frac{\pi ^2}{24 N^2}.
\end{equation}
For $N = 136$, this gives $(1-F)_{\text{gran}} < 2.2 \times 10^{-5}$, which is consistent with the oscillations in calibrated gate error that can be observed in the vertical scatter of Fig.~\ref{fig:fidelity_vs_gatetime}.

\section{Sensitivity to timing jitter}
\label{sec:jitter}

The intended suppression of the $\ket{1} \rightarrow \ket{2}$ transition depends on the relative timing of the SFQ kicks. Thus we distinguish \textit{intra-period} timing jitter, between kicks within a subsequence, from \textit{period-boundary} jitter, which is fluctuation of the boundaries between subsequences~\cite{Liebermann_2016}. The errors produced by intra-period jitter impede the geometric cancellation, whereas the errors produced by period-boundary jitter are largely in the local phase of the qubit after the gate operation. We evaluate both error channels using our previously calibrated gate as depicted in Fig.~\ref{fig:state_populations}, with $K = 4$, $\delta \theta = 0.045$ and $N = 136$. The gate time is $T_{\text{gate}} = 29.0 \text{ ns}$ and the baseline error $1 - F_0 = 4.6 \times 10^{-6}$. We do not adaptively change the already-calibrated values of $N$, $\alpha$, or our virtual $Z$-gate correction. For numerical simulation, we take the mean of 400 independent realizations per noise strength and per channel. For either noise model, the period operator is no longer identical for each period and the power in Eq.~\eqref{eq:u_total} is replaced by a time-ordered product,
\begin{equation}
    \hat{U}_{\mathrm{gate}}' = \mathcal{T}\prod_{p=0}^{N-1} \hat{U}_{\mathrm{per},p}'.
    \label{eq:u_gate_jitter}
\end{equation}

\subsection{Intra-period jitter}
\label{sec:intra_period_jitter}

Intra-period jitter is modeled by replacing the intervals between kicks given in Eq.~\eqref{eq:u_period} with new quantities $\tau_{p,k}' = \tau_K + \delta t_{p, k}$, where $p$ is the period index, $k = 1, \dots, K-1$ the interval index, and $\delta t_{p, k}$ are independent Gaussian variables with a mean of zero and standard deviation $\sigma_{\text{IPJ}}$. The start and end of each period are fixed, and so the final interval becomes
\begin{equation}
    \tau_{\mathrm{last},p}' = \tau_{\mathrm{last}} - \sum_{k=1}^{K-1} \delta t_{p,k}.
    \label{eq:ipj_last}
\end{equation}
This modifies the period operator to
\begin{multline}
    \hat{U}_{\mathrm{per},p}^{\mathrm{IPJ}} = \hat{U}_{\mathrm{free}} (\tau_{\mathrm{last},p}') \\
    \times \mathcal{T} \prod_{k=1}^{K-1} \left[ \hat{U}_{\mathrm{SFQ}} \hat{U}_{\mathrm{free}}(\tau_{p,k}') \right] \hat{U}_{\mathrm{SFQ}},
    \label{eq:u_period_ipj}
\end{multline}
where $\mathcal{T}$ is the time-ordering operator. The polygon formed by the $K$ kicks no longer closes, and these timing errors produce residual amplitudes proportional to $\delta \theta |n_{12}| \omega_{12} \delta t_{p, k}$. For independent fluctuations of this interval, the leading variance of the residual amplitude contains the geometric factor
\begin{equation}
    C_K = \frac{K}{2 \sin^2 (\pi/K)},
    \label{eq:ipj_geometric_factor}
\end{equation}
and the expectation value of the leakage attributable to this error is then, to leading order,
\begin{equation}
    \mathbb{E} [ \mathcal{L} ] - \mathcal{L}_0 \approx \frac{NC_K}{2} \left( \omega_{12} \sigma_{\text{IPJ}} \delta \theta |n_{12}| \right)^2.
    \label{eq:ipj_leakage}
\end{equation}
This approximation is not used in the full numerical calculations, which take into account the other transmon levels as well as the change in computational populations as the gate progresses.

\subsection{Period-boundary jitter}

In the period-boundary jitter model, the SFQ kicks retain their relative spacing within each subsequence, but the position of each period boundary is displaced by $\delta t_p$, an independent Gaussian variable also with zero mean and with standard deviation $\sigma_{\text{PBJ}}$. Note that both the initial and the final boundaries fluctuate. The duration of period $p$ thus becomes
\begin{equation}
    T_{\text{rep},p}' = T_{\text{rep}} + \delta t_{p+1} - \delta t_p.
\end{equation}
This modifies the final free-evolution interval
in Eq.~\eqref{eq:u_period}, giving
\begin{equation}
    \hat{U}_{\mathrm{per},p}^{\text{PBJ}} = \hat{U}_{\text{free}}(\delta t_{p+1}-\delta t_p) \hat{U}_{\text{per}}.
    \label{eq:u_period_pbj}
\end{equation}
The total evolution operator is then given by Eq.~\eqref{eq:u_gate_jitter}. Although the temporal displacement $\delta t_N-\delta t_0$ is independent of $N$, boundary fluctuations also affect the computational rotation. Neglecting the latter contribution,
we can find the leading endpoint-phase contribution to mean gate infidelity:
\begin{equation}
    \mathbb{E} [ 1 - F_{\text{avg}}] \approx ( 1 - F_0 ) + \frac{1}{3} \left( \frac{2 \pi \sigma_{\text{PBJ}}}{T_{01}} \right)^2.
    \label{eq:pbj_infidelity}
\end{equation}
This is the limit of \textit{uncorrelated} boundary-position jitter. A systematic clock error, or a clock error that varies slowly compared to the gate length, is more damaging as its phase error accumulates across the sequence, as demonstrated in Sec.~\ref{sec:detuning}.

\begin{figure}[t]
    \centering
    \includegraphics[width=\columnwidth]{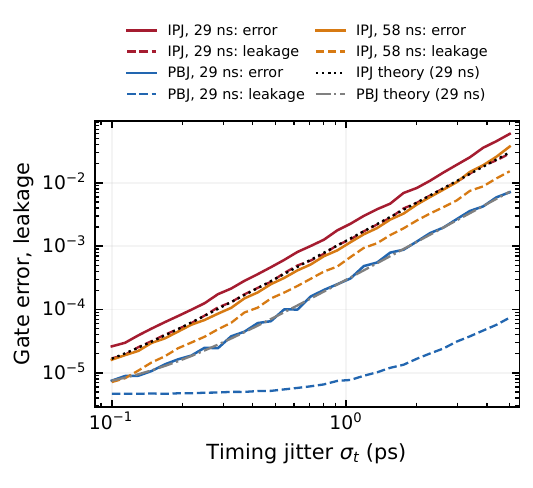}
    \caption{Average gate error (solid lines) and leakage (dashed lines) versus timing jitter for the 29-ns calibrated gate with $K = 4$ and $\delta \theta = 0.045$. Intra-period jitter breaks the geometric cancellation, whereas period-boundary jitter creates an undesirable phase in the computational subspace after the gate. Error and leakage from intra-period jitter are also shown for a 58-ns gate with $\delta\theta=0.0225$. The dotted and dash-dotted lines give Eqs.~\eqref{eq:ipj_leakage} and \eqref{eq:pbj_infidelity} respectively.}
    \label{fig:timing_jitter}
\end{figure}

Figure~\ref{fig:timing_jitter} compares the two jitter channels. Intra-period jitter is more deleterious because it breaks the geometric leakage cancellation. For our reference gate, the simulated error with intra-period jitter reaches $10^{-4}$ at just $\sigma_{\text{IPJ}} \approx 0.2 \text{ ps}$ and $10^{-3}$ with $\sigma_{\text{IPJ}}$ around 0.7 ps. Period-boundary jitter leaves leakage nearly unchanged, its effects coming from the qubit phase miscalibration, and it reaches those same error levels at $\sigma_{\text{PBJ}} \approx 0.57$ and 1.85 ps respectively. Performing a twice-longer gate by setting $\delta \theta = 0.0225$ results in less sensitivity to intra-period jitter, since the incoherent error contribution from each individual kick decreases as $\delta \theta$ decreases, but this improvement must be weighed against the increased decoherence error with a longer gate. Pulse loss, which also breaks the geometric cancellation but is unlikely with a digital pulse generator, is treated in Appendix~\ref{app:pulse_drops}.

\section{Conclusion}
\label{sec:conclusion}

We have analyzed SFQ pulse control sequences, composed of repeating subsequences in which $K$ kicks, separated by $T_{12}/K$, are repeated approximately once per computational Larmor period $T_{01}$. With our $K \geq 2$ sequences, these equally spaced kicks cancel the $\ket{1} \rightarrow \ket{2}$ transition to first order, while not canceling in the computational subspace due to anharmonicity. At zero offset charge the remaining single-period leakage into $\ket{2}$ is quartic in the kick strength, whereas leakage to higher, weakly coupled states is generally quadratic. This geometric cancellation then allows for stronger individual SFQ kicks and thus substantially shorter gates than the traditional one-kick-per-period SFQ control schemes. For the transmon considered here, calibrated $X$ sequences with $K \geq 2$ reach coherent errors below $10^{-4}$ at sampled gate times of approximately 11.5 to 14.9 ns. For $X/2$ gates, the additional freedom gained in repetition-period detuning allows the same error threshold to be reached at gate times of 4.3--6.3 ns.

For  $X$ gates, the subsequence repetition period must be detuned from the bare qubit period $T_{01}$ to compensate for the frequency shift induced by the kicks themselves, with an optimal detuning $\alpha$ proportional to the square of the kick strength. Our protocol also allows tuning based on the precise qubit spectrum: narrow resonances can arise between the harmonics of the entire sequence and higher-level transitions out of the computational subspace, and a selectable or sufficiently large value of $K$ can allow us to avoid these unfortunate coincidences.

Intra-period relative timing stability is the main remaining control requirement. Missed pulses or errors in the spacing of kicks within a single period leave a residual leakage amplitude, and uncorrelated fluctuations of the period boundaries change the computational phase in a way that cannot be simply compensated by changing the virtual $Z$ rotation at the end of the gate. Our numerical examples demonstrate that intra-period timing accuracy must be on the order of a few tenths of a picosecond to preserve coherent gate errors near $10^{-4}$, and tighter still for the lowest coherent errors shown here. This paper also does not consider additional relaxation, quasiparticle generation or heating induced by the SFQ source, which would require hardware-level calculations~\cite{Liu_2023,liu2023quasiparticle-330}.

For representative coherence times $T_1 = T_2 = 100 \, \mu \text{s}$~\cite{Bland2025MillisecondTransmons}, the faster multi-kick $X$ gates have minimal total error near $1 - F = 10^{-4}$, with decoherence being the limiting factor rather than leakage. This does not necessarily mean the establishment of operation below any particular QEC threshold, which depends on the full error model and architecture. However, it demonstrates that SFQ sequences depending on only a few parameters can substantially eliminate leakage as a dominant error source with fast single-qubit control without requiring sequences so complex that they must be generated outside of the cryogenic portion of a device.

\section*{Acknowledgements}
The authors thank Aaron Somoroff, Caleb Jordan, Alex Kirichenko, and Patrick Truitt for useful discussions. The authors also acknowledge use of the Python library Quantum Toolbox in Python (QuTiP)~\cite{Johansson2012QuTiP,Johansson2013QuTiP} in earlier stages of the numerical work.

The authors used large language models, including Claude (Anthropic, Fable 5.2 and Opus 5.5) and ChatGPT (OpenAI, GPT-6.2), during the later stages of this work. The control protocol, analytical framework, and original simulation code were developed by the authors. Under the authors' direction, generative AI assisted with verification of analytical results, refinement and execution of numerical code, preparation of figures (including the code used to render all figures) and manuscript editing. The authors reviewed and verified the results and take full responsibility for the content of this article.

\appendix

\section{Detuning of the qubit period}
\label{app:alpha}

\begin{figure}[t]
    \centering
    \includegraphics[width=\columnwidth]{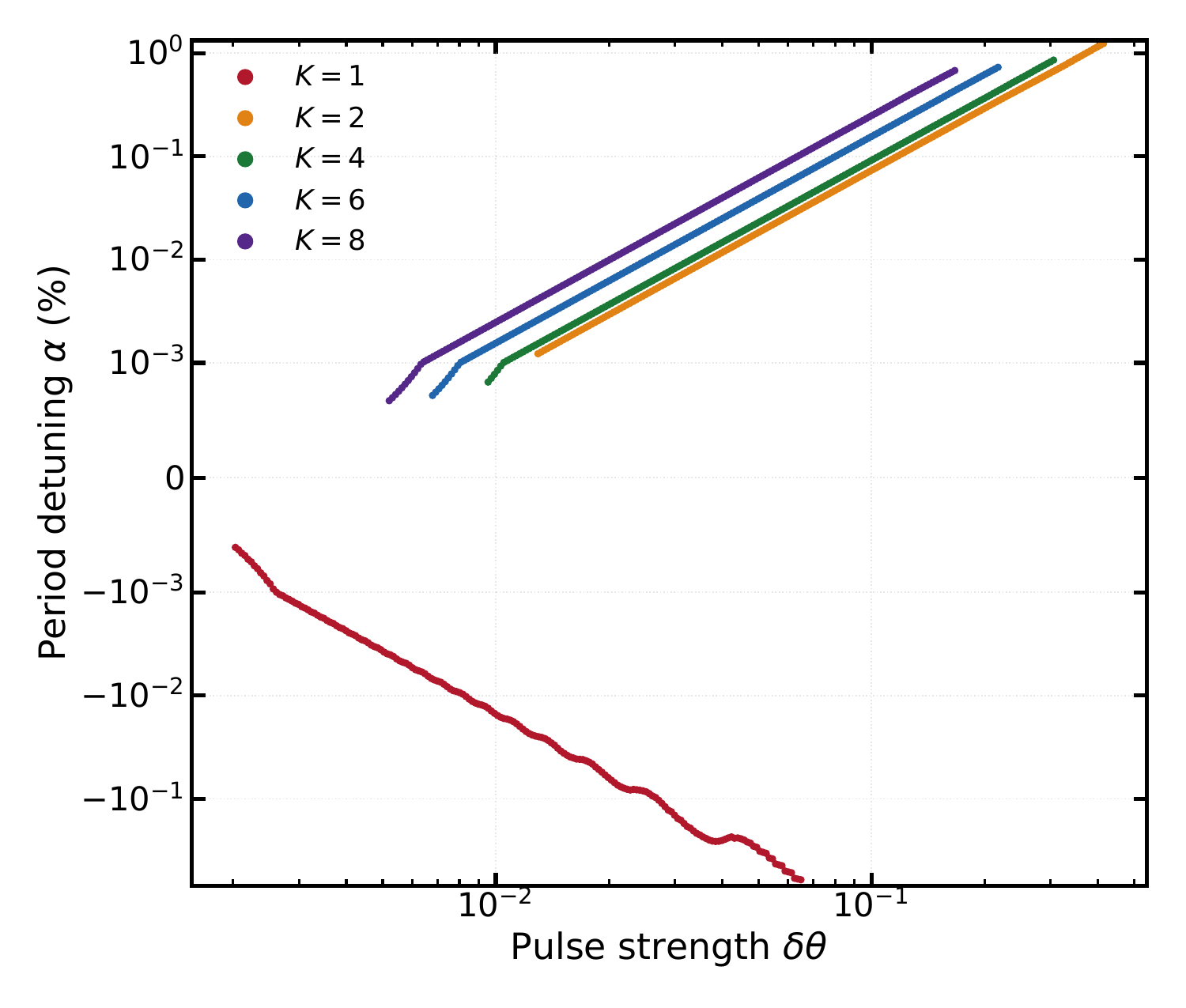}
    \caption{Optimization of the repetition period detuning $\alpha$ as a function of kick strength $\delta \theta$ for $K \in \{ 1, 2, 4, 6, 8 \}$. At weak kick strengths the detuning is quadratic in $\delta \theta$, and the lines represent $\alpha = c_K \delta \theta^2$, with $c_1=-1.49$, $c_2=0.073$, $c_4=0.092$, $c_6=0.155$, and $c_8=0.247$. These coefficients are fitted to the calibrated points with $\delta\theta\leq0.05$ ($\delta \theta \leq 0.03$ for $K = 1$, the curve of which is not a clean quadratic). The vertical axis is in percent. This compensates for the shift in the qubit Larmor period induced by the SFQ kicks themselves.}
    \label{fig:alpha}
\end{figure}

The repetition period detuning $\alpha$ compensates for the effect of the SFQ sequence itself on the qubit phase. Here we consider sequences with $K \geq 2$. In the interaction picture, with $\hat{H}_{\text{st}}$ as the static transmon Hamiltonian, the kick at time $t_k$ is generated by the operator $\hat{n}_I (t_k) = e^{i \hat{H}_{\text{st}} t_k / \hbar} \, \hat{n} \, e^{-i \hat{H}_{\text{st}} t_k / \hbar}$. The Magnus expansion~\cite{Ribeiro2017MagnusLeakage,Choi2017QuditEngineering,Zhou2024QuditEngineering} of the propagator for one period contains the commutator term
\begin{equation}
    \Omega_2=-\frac{\delta\theta^2}{2} \sum_{k>l}[\hat n_I(t_k),\hat n_I(t_l)]
    \label{eq:magnus_omega2}
\end{equation}
to second order in $\delta \theta$. Its diagonal elements effect a computational relative phase shift. Specifically, we can insert a complete set of energy eigenstates between the operators in the commutator and recover
\begin{equation}
   \bra{a} [\hat n_I(t_k),\hat n_I(t_l)]\ket{a} =-2i\sum_c |n_{ac}|^2 \sin[\omega_{ac}(t_k-t_l)],
    \label{eq:charge_commutator}
\end{equation}
with $\omega_{ac} = ( E_c - E_a) / \hbar$. The sum over $c$ accounts for intermediate states that occur in second-order processes which return the transmon to $\ket{a}$. Noting that $t_k - t_l = (k - l) \tau_K$, we can substitute Eq.~\eqref{eq:charge_commutator} into Eq.~\eqref{eq:magnus_omega2} and get
\begin{equation}
    \bra{a}\Omega_2\ket{a} = i\delta\theta^2 \sum_c |n_{ac}|^2 \sum_{k>l} \sin[\omega_{ac}\tau_K(k-l)].
\end{equation}
Since there are $K-d$ pairs of kicks separated by $d = k - l$ intervals, we can rewrite this as a single sum:
\begin{equation}
    \sum_{k>l}\sin[\omega_{ac}(k-l)\tau_K] = \sum_{d=1}^{K-1}(K-d) \sin(\omega_{ac}\tau_K d).
\end{equation}
For convenience, we define $\varphi = \omega_{ac} \tau_K $ as the phase accumulated between two successive kicks, and express the information from the above equation with the function
\begin{equation}
    \Lambda_K(\varphi) = \sum_{d=1}^{K-1}(K-d)\sin(d \varphi).
    \label{eq:lambda_k}
\end{equation}
The diagonal elements of $\Omega_2$ can then be expressed as
\begin{equation}
    \bra{a} \Omega_2 \ket{a} = i \delta \theta^2 \sum_c |n_{ac} |^2 \Lambda_K (\omega_{ac} \tau_K ).
\end{equation}
Noting that $\Omega_2$ is anti-Hermitian, we can write that $\bra{a} \Omega_2 \ket{a} = i \chi_a$. Then $\chi_a$ is real and represents the second-order phase contribution from state $\ket{a}$. We can then obtain
\begin{equation}
    \begin{aligned}
    \chi_1 - \chi_0 &= \Delta \phi_{\text{SFQ}} \\
    & \simeq \delta \theta^2 \left[ |n_{12}|^2 \Lambda_K (\omega_{12} \tau_K) - 2 |n_{01}|^2 \Lambda_K (\omega_{01} \tau_K ) \right].
    \end{aligned}
    \label{eq:sfq_phase_shift}
\end{equation}
The first term in $\Delta \phi_{\text{SFQ}}$ comes from virtual transitions between $\ket{1} $ and $\ket{2}$, whereas the second term is from the composition of successive rotations occurring about different axes within the computational subspace. It should be noted that both of these terms are substantially larger than their difference for $K \geq 3$.

Since $\tau_K=T_{12}/K$, we have $\omega_{12}\tau_K=2\pi/K$. Using Eq.~\eqref{eq:lambda_k}, we find
\begin{equation}
    \Lambda_K \left( \frac{2\pi}{K} \right) =\frac{K}{2} \cot \left( \frac{\pi}{K} \right).
\end{equation}
By way of compensation for the phase shift from Eq.~\eqref{eq:sfq_phase_shift}, we set the subsequence repetition time $T_{\text{rep}} = (1 + \alpha) T_{01}$, thus introducing a relative phase of $-2 \pi \alpha$. Thus we cancel the relative phase shifting by setting
\begin{equation}
    \alpha = \frac{1}{2 \pi} \Delta \phi_{\text{SFQ}}.
    \label{eq:alphafinal}
\end{equation}
Equivalently, the effective \textit{dressed} computational frequency associated with this phase shift is
\begin{equation}
    \tilde{\omega}_{01} \simeq \omega_{01} - \frac{\Delta\phi_{\mathrm{SFQ}}}{T_{\mathrm{rep}}}.
    \label{eq:dressed_omega01}
\end{equation}
For the calibrated $X$ gate, $\tilde{\omega}_{01} T_{\mathrm{rep}} \simeq 2 \pi$. With fixed transmon parameters, this gives $\alpha \simeq c_K \delta \theta^2$. Evaluating Eq.~\eqref{eq:alphafinal} with the two terms of Eq.~\eqref{eq:sfq_phase_shift} gives $c_K = 0.073, 0.092, 0.155$ and 0.246 for $K = 2, 4, 6$ and 8, within 1\% of the fitted values, with no adjustable parameter.

\section{Finite pulse duration}
\label{app:width}

The delta-function approximation in Eq.~\eqref{eq:u_sfq} treats each SFQ pulse as having zero duration. We have verified its accuracy by replacing the delta-function kick with a Gaussian voltage pulse with the same integrated kick strength $\delta \theta$ and a full duration at half maximum of 1--12 ps for $K=4$ and 1--4 ps for $K=8$, integrated numerically using 48 sub-steps of $\exp [ -i (\hat{H}/\hbar - \delta \theta g(t) \hat{n} ) dt ]$, over a duration of $\pm 4$ rms widths around each pulse center. (Here $\int g \, dt = 1$.) For the calibrated gate of Sec.~\ref{sec:calibration}, with $K = 4, \, \delta \theta =0.045$; a 2 ps pulse changes the leakage by $0.6 \%$ and the coherent gate error from $4.6 \times 10^{-6}$ to $5.8 \times 10^{-6}$ before recalibration of $\alpha$. Recalibration then gives us a gate error of $4.7 \times 10^{-6}$.

To first order in the kick strength, finite pulse width filters the transition amplitude at a given frequency $f$ by the Fourier transform of the pulse. Taking $\sigma$ to represent the rms pulse duration, this is equal to $\exp ( -2 \pi^2 f^2 \sigma^2 )$. With a 2 ps pulse, this value is equal to 0.9997 at $f_{01}$ and 0.998 at $f_{14}$. As this modifies the effective computational rotation per subsequence $\Theta$ and the calibrated value of $\alpha$, gates recalibrated with finite pulse duration taken into account (allowing adjustment of $N$, $\alpha$ and the final virtual-$Z$ phase) retain error at the $10^{-5}$ level for widths up to 10 ps at $K = 4$ ($\tau_K = 56$ ps, shortest interval 45 ps) and up to 4 ps at $K=8$ ($\tau_K = 28$ ps, shortest interval 17 ps). The leakage decreases slightly as the filter attenuates the kick at higher transition frequencies. Intuitively, without recalibration, the slight attenuation of the effective kick at $f_{01}$ leads to under-rotation.

\section{Resonant leakage to higher transmon levels}
\label{app:res16}

\subsection{Coupling matrix elements and resonance conditions}

As discussed in Sec.~\ref{sec:perturbation}, for some values of $K$ and $\tau_K$, there can be leakage out of the computational subspace due to resonance with higher-level transitions, i.e., a higher harmonic of the full sequence may be resonant with $f_{16}$, as in Fig.~\ref{fig:fidelity_vs_ng}. The leakage from these channels is generally low due to the small magnitude of the charge matrix element coupling higher levels with computational levels, but can be significant in the case of a particularly strong resonance~\cite{Dumas2024TransmonIonization}. We use the $\ket{1} \rightarrow \ket{6}$ resonance from Fig.~\ref{fig:fidelity_vs_ng} as our worked example throughout this appendix.

\subsubsection{Matrix element.}

In the harmonic expansion of a transmon, we have $\hat{n} = i n_0(\hat{a}^\dagger - \hat{a})$, with $n_0 = (E_J / 32 E_C)^{1/4}$ as in Sec.~\ref{sec:perturbation}; the charge operator increments or decrements the oscillator quantum number by one, and the process represented by $\matrixel{1}{\hat{n}}{6}$ requires mixing either four or six quanta from the anharmonic portion of the transmon Hamiltonian represented by $-E_J \cos \hat{\varphi}$. The first-order contribution of the quartic term vanishes, as the paths through $\ket{2}$ and $\ket{5}$ have equal and opposite weight. As such, the leading term is of second order in the anharmonicity and $|n_{16}|^{(2)} \propto (E_C / E_J )^{3/4}$. Exact diagonalization approaches this limit only when $E_J / E_C \gtrsim 200$. For $E_J/E_C = 75, 60 \text{ and } 50$ we find $|n_{16}| = 0.0063, 0.0085 \text{ and } 0.0120$ respectively at $n_g = 0$, which exceed the perturbative prediction by factors of $1.5, 1.7 \text{ and }$ 2.0. This has the same origin as the large charge dispersion of $\ket{6}$: its energy lies within 3.5 GHz of the top of the transmon barrier, where the wave function is spreading over the well and the oscillator expansion is no longer sufficient. The coupling is thus small compared with $|n_{12}| \approx 1.57$ (at $E_J/E_C = 60$; 1.67 at 75), but not exponentially so.

\subsubsection{Peak height.}

To first order in $\delta \theta$, the kicks from a single period generate an amplitude $\delta \theta |n_{16} | |S_K (\omega_{16})|$ from the $\ket{1}$ component of the statevector. The population of $\ket{1}$ is not constant throughout any gate effecting a rotation between $\ket{0}$ and $\ket{1}$. At exact resonance, after $N$ periods, the leaked population becomes
\begin{equation}
    P_6 (N) \simeq \left( \frac{2N}{\pi}  \delta \theta |n_{16}||S_K (\omega_{16}) | \right) ^2,
    \label{eq:p6_again}
\end{equation}
to leading order for either computational basis input. Note that the factor $2/ \pi$ comes from summing the computational-state amplitude as it changes over the $\pi$-rotation:
\begin{equation}
    \sum_{p=1}^N \sin \left( \frac{\pi p}{2N} \right) \approx \frac{2N}{\pi}.
\end{equation}
When this leakage dominates, $1 - F_{\text{avg}} \approx P_6$, averaged over both computational basis inputs. For our gate of interest from Fig.~\ref{fig:fidelity_vs_ng}, $|n_{16} | = 0.0079, |S_4 (\omega_{16} ) | = 3.6$ at the \textit{resonant} offset charge $|n_g| \approx 0.35$, and Eq.~\eqref{eq:p6_again} gives $P_6 = 1.0 \times 10^{-2}$ while the numerical calculation gives an average gate infidelity of $1.16 \times 10^{-2}$.

\subsubsection{Identifying resonant channels.}

Two conditions must be satisfied before leakage into a higher transmon level becomes significant. First, the transition must not be suppressed within a subsequence, which requires the intra-period interference or comb factor $S_K (\omega_{1n})$ of Eq.~\eqref{eq:comb} to be sufficiently large. Second, the dressed transition frequency $\tilde{f}_{1n}$ must be sufficiently close to a harmonic of the repetition frequency that the residual amplitudes from the subsequences add coherently:
\begin{equation}
    \tilde{f}_{1n} T_{\text{rep}} \approx m, \qquad m \in \mathbb{Z}.
    \label{eq:dressed_resonance}
\end{equation}
Note that the two resonance conditions do not necessarily need to have the same harmonic index. Then, as per Eq.~\eqref{eq:p6_again}, the strength of the resonance is determined by the magnitude of the charge matrix element $|n_{1n}|$. The strongest resonances are to be expected when $f_{1n} \approx m K f_{12}$ for some integer $m$. This first becomes the case with the $\ket{1} \rightarrow \ket{6} $ transition in the worked example used in this appendix. Other transitions may satisfy one or the other of those two conditions, but both are required to have substantial resonance.

\subsubsection{Position of the resonance.}

The resonance condition from Eq.~\eqref{eq:dressed_resonance} involves the \textit{dressed} transition frequency. As discussed in Appendix~\ref{app:alpha}, the application of the SFQ kicks modifies the transmon spectrum. For $E_C / h = 0.25 \text{ GHz}$, the bare frequency collision $f_{16} T_{\text{rep}} = 4$ occurs at $|n_g| \approx 0.30$, whereas the leakage peak in Fig.~\ref{fig:fidelity_vs_ng} occurs at $|n_g| \approx 0.35$.

There are two contributions to the displacement. The first is the time-averaged gate voltage associated with the SFQ sequence, which produces an effective offset charge of magnitude $\hbar K \delta \theta / 8 E_C T_{\text{rep}}$ (Eq.~\eqref{eq:delta_n_g}). This leads to a shift in the transmon energy levels, which can be elucidated with the Hellmann--Feynman theorem:
\begin{equation}
    \bra{i} \hat{n} \ket{i} - n_g = -\frac{1}{8 E_C} \frac{\partial E_i}{\partial n_g}.
\end{equation}
Since the charge dispersion is much greater for the higher transmon levels, this effective offset charge substantially shifts the higher transition frequencies while effecting a comparatively small change in the qubit frequency. It also largely accounts for the displacement of the symmetry of the leakage curves away from $n_g = 0$.

The second contribution is due to the kicks producing frequency shifts via virtual transitions between the transmon levels analogous to those considered in Appendix~\ref{app:alpha}. These further displace the resonance condition. Both contributions are included in the quasienergy spectrum of $\hat{U}_{\text{per}}$, which provides a more precise description of the resonance than the bare transmon spectrum. However, away from these higher-level resonances, the phase mismatch between successive subsequences prevents coherent accumulation of leakage amplitudes.

\subsection{Suppression of higher-level resonances}

The dependence of the intra-period interference factor on the number of kicks provides a method of suppressing these resonances without changing the qubit characteristics. In particular,
\begin{equation}
    S_{2K} (\omega) = \left( 1 + e^{i \omega \tau_K / 2} \right) S_K (\omega).
    \label{eq:doublecomb}
\end{equation}
Thus, as discussed in the body of the paper, doubling the number of kicks zeroes the odd principal maxima of the original sequence. Fig.~\ref{fig:fidelity_vs_ng} demonstrates the suppression of one of these resonances by doubling $K$. Additionally, for sufficiently large numbers of kicks, the first principal maximum of $S_K$ lies outside the transmon potential's energy barrier. The charge matrix elements for these states are small, and so high-$K$ sequences generally cause few problems with higher-level resonances. $K$ can thus be chosen in reference to the qubit parameters: for the transmons examined in this work, $K = 4$ is sufficient when $E_J / E_C = 75$, whereas $K = 8$ is generally preferred when $E_J / E_C$ is lower.

\begin{figure}[t]
    \centering
    \includegraphics[width=\columnwidth]{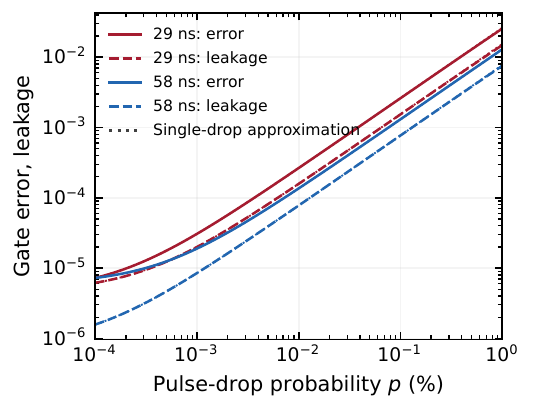}
    \caption{Mean gate infidelity (solid) and leakage (dashed) versus the independent per-kick pulse-drop probability $p$ for calibrated 29-ns and 58-ns gates. The average uses the exact independent-drop channel while retaining the original virtual-$Z$ correction. Dotted curves give the single-drop approximation $\mathbb{E}[M(p)] \simeq M_0 + pKN \overline{\Delta M}$, with $\Delta M$ representing change in infidelity or leakage as appropriate and $\overline{\Delta M}$ representing its mean. The horizontal axis shows $p$ in percent.}
    \label{fig:pulse_drops}
\end{figure}

\section{Stochastic pulse drops}
\label{app:pulse_drops}

Imperfections in the SFQ control hardware can create stochastic pulse drops. We analyze this channel because a single missed kick disrupts the geometric cancellation. Our model is of a stochastic probability $p$ for any given kick to be dropped, applied to the same calibrated gate as discussed in Sec.~\ref{sec:jitter}. If a pulse is dropped, the corresponding $\hat{U}_{\text{SFQ}}$ is replaced with the identity.

Mathematically, we let $d_{n, m} \in \{0, 1\}$ be a Bernoulli random variable corresponding to the $m$th pulse in the $n$th period, equal to 1 with probability $p$. When $d_{n,m} = 1$ the pulse is dropped. The period operator for the $n$th period is then recomposed with the identity replacing $\hat{U}_{\text{SFQ}}$ for each dropped pulse. A dropped kick then leads to a residual amplitude of $\delta \theta |n_{12}|$ due to removing one side of the geometric cancellation $K$-gon. With an average of $pKN$ drops per gate, that means the mean leakage is
\begin{equation}
    \mathbb{E} [\mathcal{L}_{\text{drop}} ] \simeq \mathcal{L}_0 + \frac{1}{2} pKN (\delta \theta |n_{12} | )^2,
    \label{eq:l_drop}
\end{equation}
when considering only the $\ket{1}$ component, as $\ket{1} \rightarrow \ket{2}$ leakage is the overwhelmingly dominant channel when geometric cancellation fails. For the reference gate, deterministic enumeration of every possible single-pulse loss gives a mean leakage increase of $2.8226\times10^{-3}$ per lost pulse, compared with $2.8244\times10^{-3}$ from Eq.~\eqref{eq:l_drop}. As Monte Carlo sampling is difficult at extremely small pulse drop probabilities, the numerical average is evaluated by considering the independent-drop channel $\mathcal{E}_p(\rho)=(1-p)\hat{U}_{\mathrm{SFQ}}\rho\hat{U}_{\mathrm{SFQ}}^\dagger+p\rho$ along with the free-evolution channels.

Figure~\ref{fig:pulse_drops} demonstrates the impact of pulse drops on gate performance. For the calibrated 29-ns gate with $KN=544$, the mean infidelity reaches $10^{-4}$ at $p\simeq3.66\times10^{-5}$ when the virtual-$Z$ correction is held fixed, as appropriate for unobserved pulse loss. For the approximately 58-ns gate, the corresponding threshold is $p\simeq7.23\times10^{-5}$. Per event, pulse loss is the most catastrophic hardware failure considered, although it is expected to be rare in a well-engineered system.

\section{Extension to $X/2$ gates}
\label{app:xhalf}

An $X/2$ gate can be implemented via a virtual-$Z$ rotation, a physical rotation about an axis tilted outside the equatorial plane, and a second virtual-$Z$ rotation. Let $R_{\beta}(\gamma)$ denote a rotation through an angle $\gamma$ about an axis tilted off the equator by an angle $\beta$. Because the frame rotations do not change the magnitude of any matrix elements, $Z(\phi_f) R_\beta (\gamma) Z ( \phi_i )$ effects an $X/2$ gate up to an overall phase provided that
\begin{equation}
    \cos^2 \beta \, \sin^2 (\gamma/2) = \frac{1}{2}.
\end{equation}

It should be noted that reprogramming $\alpha$ requires adjustment of the subsequence repetition period, whereas the virtual-$Z$ gates can be implemented as frame updates.

For our numerical calculations, we let $\hat{U}_{\text{seq, }2 \times 2}$ denote the computational block of the quasi-gate operator $\hat{U}_{\text{per}}^N$, without the virtual-$Z$ frames, and we let $\hat{U}_{\text{gate, } 2 \times 2}$ denote the full gate operator:
\begin{equation}
    \hat{U}_{\text{gate, } 2 \times 2} = Z (\phi_f ) \hat{U}_{\text{seq, }2 \times 2} Z( \phi_i).
\end{equation}
Then the fidelity $F_{X/2}$, inclusive of leakage, can be represented as
\begin{equation}
\begin{aligned}
F_{X/2} = \frac{1}{6}\Bigl( &\left|\mathrm{Tr}\left[ R_x^\dagger(\pi/2) \hat{U}_{\text{gate},2\times2} \right]\right|^2 \\
&+ \mathrm{Tr}\left( \hat{U}_{\text{seq},2\times2}^\dagger \hat{U}_{\text{seq},2\times2} \right) \Bigr).
\end{aligned}
\end{equation}
As we wish to find $F_{\text{max, } X/2} = \max_{\phi_i, \phi_f} F_{X/2}$, we employ the following trick to reduce it to a one-dimensional search. We write $U_{\text{seq, } 2 \times 2}=\bigl ( \begin{smallmatrix} a&b\\ c&d \end{smallmatrix} \bigr)$ and $\mathcal{T} = \mathrm{Tr} [ R_x^\dagger (\pi/2) \hat{U}_{\text{gate, } 2 \times 2} ]$, and then we have
\begin{equation}
  \begin{aligned}
  \max_{\phi_i, \phi_f} | \mathcal{T} | &=\frac{1}{\sqrt2}\max_{0 \leq \chi < \pi} \\
  &\quad\left(|a+e^{2 i \chi}d^* | + | b-e^{2 i \chi} c^* | \right).
  \end{aligned}
  \label{eq:xhalf_exact_phase}
\end{equation}
This is valid for any nonunitary computational block. For each sampled kick strength, our numerical optimization selects the value of $N$ on the first-rotation branch and optimizes $\alpha$ within the approximate detuning window. The virtual-$Z$ phases $\phi_i, \phi_f$ are then optimized using Eq.~\eqref{eq:xhalf_exact_phase}. Both sweeps in Fig.~\ref{fig:fidelity_vs_gatetime} use 20 transmon eigenstates, and the calibrated gates are re-evaluated at 25 levels without making changes to their parameters or virtual-$Z$ phases. The complete sweep and gate parameters are provided in the reproduction code.

For a representative example, the fastest sampled $K = 4$ gate with coherent error less than $10^{-4}$ has $\delta \theta = 0.15193$, $\alpha = -0.009975$, $N = 30$ and $T_{\text{gate}} = 6.33$ ns for a coherent error of $1 - F = 9.27 \times 10^{-5}$. The substantially tilted physical rotation in this case is allowed by the additional freedom in $\alpha$. The shortest sampled $K = 8$ gate below our $10^{-4}$ coherent error threshold has a much smaller axis tilt, and so its speedup is largely accomplished via a reduced logical rotation angle.

Although $R_x(\pi/2)^2$ effects an $X$ gate up to an overall phase, composition of an $X$ gate in this manner requires care in implementing the virtual-$Z$ rotations, as residual leakage amplitudes can evolve and interfere during the second operation. The infidelity of the composite $X$ gate is not determined solely by the infidelity of the $X/2$ gate, and this composition does not generally produce better results than the typical $X$ gate described in this paper. As such this composite control is not included in Fig.~\ref{fig:fidelity_vs_gatetime}.

\bibliography{SFQ_bibliography_complete}

\end{document}